\documentclass{aa}  
\usepackage{graphicx}
\usepackage{amsmath}
\usepackage{amssymb}
\usepackage{txfonts}

\newcommand{\ros}{{\it ROSAT}}

\newcommand{\xmm}{{\it XMM-Newton}}

\newcommand{\eros}{{eROSITA}}

\newcommand{\srgl}{{\it Spectrum Roentgen Gamma (SRG)}}
\newcommand{\srg}{{\it SRG}}

\newcommand{\cxo}{{\it Chandra}}

\newcommand{\esol}{{European Southern Observatory Very Large Telescope}}
\newcommand{\eso}{{\it ESO-VLT}}

\newcommand{\nh}{N_{\rm H}}
\newcommand{\gaia}{{\it GAIA}}
\newcommand{\forstl}{{Focal Reducer/low dispersion Spectrograph 2}}
\newcommand{\forst}{{FORS2}}
\newcommand{\fastl}{{Five-hundred-meter Aperture Spherical radio Telescope}}
\newcommand{\fast}{{\it FAST}}

\def \tmzsfs{\object{PSR~B0656+14}}
\def \onon{\object{PSR~J1909--3744}}

\def \fluxcgs{erg~s$^{-1}$~cm$^{-2}$}
\def \jotos{\object{eRASSU~J131716.9--402647}}
\def \jzsfs{\object{eRASSU~J065715.3+260428}}

\def \jostt{\object{eRASSU~J072302.3--194225}}
\def \joeon{\object{eRASSU~J081952.1--131930}}
\def \joefo{\object{eRASSU~J084046.2--115222}}
\def \jotfs{\object{eRASSU~J134725.4--363415}}
\def \jofff{\object{eRASSU~J144516.0--374428}}

\def \ostt{\object{J0723}}
\def \oeon{\object{J0819}}
\def \oefo{\object{J0840}}
\def \otfs{\object{J1347}}
\def \offf{\object{J1445}}

\def \magoe{\object{RX~J1856.5--3754}}

\def \magzf{\object{RX~J0420.0--5022}}

\begin{document} 
\title{
X-ray, optical, and radio follow-up of five thermally emitting isolated neutron star candidates\thanks{Based on observations obtained with \xmm, an ESA science mission with instruments and contributions directly funded by ESA Member States and NASA (observations 0921280301, 0921280401, 0921280501, 0921280601, 0921280701)}}
\author{J.~Kurpas\inst{1}
\and A.~M.~Pires\inst{2,1}
\and A.~D.~Schwope\inst{1}
\and B.~Li\inst{3}
\and D.~Yin\inst{3}
\and F.~Haberl\inst{4}
\and M.~Krumpe\inst{1}
\and S.~Sheth\inst{1,5}
\and I.~Traulsen\inst{1}
\and Z.~L.~Zhang\inst{6,7}
}
\offprints{J. Kurpas}
\institute{Leibniz-Institut f\"ur Astrophysik Potsdam (AIP), An der Sternwarte 16, 14482 Potsdam, Germany
\email{jkurpas@aip.de} 
\and
Center for Lunar and Planetary Sciences, Institute of Geochemistry, Chinese Academy of Sciences, 99 West Lincheng Rd., 550051 Guiyang, China
\and
College of Physics, Guizhou University, Guiyang 550025, China
\and
Max-Planck-Institut f\"ur extraterrestrische Physik, Gie{\ss}enbachstra\ss e 1, 85748 Garching, Germany
\and
Potsdam University, Institute for Physics and Astronomy, Karl-Liebknecht-Stra\ss e 24/25, 14476 Potsdam, Germany
\and
Shanghai Astronomical Observatory, Chinese Academy of Sciences, Shanghai 200030, China
\and
Key Laboratory of Radio Astronomy and Technology, Chinese Academy of Sciences, Beijing, People’s Republic of China
}
\date{Received ...; accepted ...}
\keywords{pulsars: general --
stars: neutron -- X-rays: general}
\titlerunning{Follow-up of five INS candidates}
\authorrunning{J.~Kurpas et al.}
\abstract
{
We report on follow-up observations with \xmm, the \forst\ instrument at the \eso, and FAST, aiming to characterise the nature of five thermally emitting isolated neutron star (INS) candidates recently discovered from searches in the footprint of the \srgl/\eros\ All-sky Survey. We find that the X-ray spectra are predominantly thermal and can be described by low-absorbed blackbody models with effective temperatures ranging from 50 to 210~eV. In two sources, the spectra also show narrow absorption features at $300 - 400$~eV. Additional non-thermal emission components are not detected in any of the five candidates. The soft X-ray emission, the absence of optical counterparts in four sources, and the consequent large X-ray-to-optical flux ratios $>3000 - 5400$ confirm their INS nature. For the remaining source, \jofff, the available data do not allow a confident exclusion of an active galactic nucleus nature. However, if the source is Galactic, the small inferred X-ray emitting region is reminiscent of a heated pulsar polar cap, possibly pointing to a binary pulsar nature. X-ray timing searches do not detect significant modulations in all candidates, implying pulsed fraction upper limits of 13 -- 19\% ($0.001-13.5$~Hz). The absence of pulsations in the FAST observations targeting \joeon\ and \joefo\ excludes periodic magnetospheric emission at 1 -- 1.5~GHz with an $8\sigma$ significance down to 4.08~µJy and 2.72~µJy, respectively. The long-term X-ray emission of all sources does not imply significant variability. Additional observations are warranted to establish exact neutron star types. At the same time, the confirmation of the predominantly thermal neutron star nature in four additional sources highlights the power of \srg/\eros\ to complement the Galactic INS population.
}
\maketitle
\section{Introduction\label{sec_intro}
}


Isolated neutron star (INS) searches at X-ray energies are an important tool to complement the known population. This is because they not only are susceptible to the magnetospheric emission or accretion processes that drive INS discoveries at other wavelengths \citep[e.g. in the radio or gamma-ray regime;][]{2025RAA....25a4001H,2023ApJ...958..191S}, but also are sensitive to thermal emission components originating directly from the INS surface. This particularly benefits the discovery of still rare predominantly thermally emitting INS types, such as the X-ray dim INSs \citep[XDINSs;][]{2009ASSL..357..141T}, or radio and gamma-ray-quiet rotation-powered pulsars (RPPs) with an unfavourable magnetospheric viewing geometry that are otherwise missed in conventional pulsar searches. Likewise, a growing population of thermally emitting INSs of all types ought to improve studies targeting the state and composition of neutron star atmospheres and matter \citep{2002nsps.conf..263Z,2016ARA&A..54..401O}, the phenomenology of the known INS population and its magneto-thermal evolution \citep[e.g.][]{2013MNRAS.434..123V,2021ApJ...914..118D}, and the calibration of INS emission and cooling models \citep[e.g.][]{2020MNRAS.496.5052P}. Consequently, the discovery and characterisation of new INSs at X-ray energies is of high interest.

Between December 2019 and February 2022, the \eros\ instrument aboard the \srgl\ mission conducted the \srg/eROSITA All-Sky Survey \citep[eRASS;][]{2021A&A...647A...1P, 2024A&A...682A..34M}. In total, 4.3 all-sky surveys were performed in the soft X-ray band (0.2 -- 5~keV). With a much improved sensitivity and astrometric precision in relation to previous all-sky surveys at X-ray energies \citep[as were exemplarily conducted with ROSAT;][]{1982AdSpR...2d.241T} and the wide survey area much exceeding the sky covered in serendipitous source catalogues from pointed X-ray missions such as \xmm\ \citep{2020A&A...641A.136W} and \cxo\ \citep{2024ApJS..274...22E}, the eRASS is ideal for identifying new predominantly thermally emitting INSs. Recently, a search in the eRASS data covering the western Galactic hemisphere down to an X-ray flux limit of $10^{-13}$~\fluxcgs\ allowed the collection of a sample of $\sim30$ promising candidates \citep{2024A&A...687A.251K}. Due to short exposures and the specific survey pattern, the eRASS data alone do not allow for a detailed spectral or timing study of the selected sources. Similarly, the archival photometric catalogues used to screen for soft X-ray emitting contaminants with low X-ray-to-optical flux ratios, such as cataclysmic variables (CVs) and active galactic nuclei (AGNs), are not sufficiently deep to exclude a non-INS nature for any of the selected candidates. Consequently, follow-up observations are required to fully establish the nature of the sources. Such observations were recently carried out in a large programme at X-ray and optical wavelengths for seven of the most promising candidates with \xmm\ \citep{2001A&A...365L...1J} and the \forstl\ (\forst) at the \esol\ \citep[\eso;][]{1998Msngr..94....1A}. The results of this observational campaign on two sources, \jzsfs\ and \jotos, have already been published, leading to the identification of a thermally emitting, but radio and gamma-ray faint, RPP \citep[][]{2025A&A...694A.160K} and the discovery of a highly magnetised long-period ($P\sim12.8$~s) INS resembling the known XDINSs \citep{2024A&A...683A.164K}. Building on these exciting discoveries, we present the observational results and discuss the nature of the remaining five targets, \jostt, \joeon, \joefo, \jotfs, and \jofff\ (henceforth dubbed \ostt, \oeon, \oefo, \otfs, and \offf), in this work. For two of them (\oeon and \oefo), the observational coverage is further extended with radio data from the \fastl\ \citep[\fast,][]{2011IJMPD..20..989N}.

This paper is outlined as follows. In Sect.~\ref{sec_obs} we describe the general data reduction, and in Sect.~\ref{sec_analysis} we present our results. Finally, we discuss the nature of the candidates and present our conclusions in Sect.~\ref{sec_disc} and Sect.~\ref{sec_concl}.

\begin{table}
\small
\caption{Summary of follow-up observations \label{tab_obs}}
\centering
\scalebox{.84}{
\begin{tabular}{lccccc}
\hline\hline\noalign{\smallskip}
\multicolumn{5}{l}{\xmm\tablefootmark{(a)}}\\
\hline
Source & MJD\tablefootmark{(b)} & Counts\tablefootmark{(c)} & Background & Net count & \multicolumn{1}{c}{GTI\tablefootmark{(d)}}\\
 & & & contribution\tablefootmark{(c)} & rate\tablefootmark{(c)} & \\
& [days] & & [\%] &[$10^{-2}$\,cts\,s$^{-1}$] & \multicolumn{1}{c}{[s]} \\
\hline
\ostt & 60402.37473 & 4135 & $8.3 \pm 1.7$  & 8.20(15)  & 51329 \\
\oeon & 60073.48576 & 6892 & $10.0 \pm 1.3$ & 13.10(19) & 54134 \\
\oefo & 60237.30717 & 5413 & $7.3 \pm 1.5$ & 12.89(20) & 43349 \\
\otfs & 60163.26739 & 2919 & $12.5\pm 2.0$ & 6.63(15)  & 42953 \\
\offf & 60185.36513 & 3193 & $8.7 \pm 1.9$ & 8.93(18)  & 36396 \\
\hline
\multicolumn{5}{l}{\eso}\\
\hline
Source & MJD\tablefootmark{(b)} & & & & \multicolumn{1}{c}{T$_\mathrm{exp}$} \\
& [days] & & & & \multicolumn{1}{c}{[s]}\\
\hline
\ostt & 60254.27166 & & & & 4400 \\
\oeon & 60076.98230 & & & & 3200 \\
\oefo & 60077.98398 & & & & 3200 \\
\otfs & 60082.13879 & & & & 3600 \\
\offf & 60078.23744 & & & & 4000 \\
\hline
\multicolumn{5}{l}{\fast}\\
\hline
Source & MJD\tablefootmark{(b)} & & & & \multicolumn{1}{c}{T$_\mathrm{exp}$} \\
& [days] & & & & \multicolumn{1}{c}{[s]}\\
\hline
\oeon & 60587.34583 & & & & 1200 \\
\oefo & 60559.43854 & & & & 2700 \\
\hline
\end{tabular}}
\tablefoot{
\tablefoottext{a}{Values for EPIC-pn only.}
\tablefoottext{b}{Modified Julian date at mid-observation.}
\tablefoottext{c}{Value given for the energy band of $0.2-12$\,keV.}
\tablefoottext{d}{Remaining ‘good’ observing time intervals with data reduction steps applied and periods of high background activity screened (see text for details).}
}
\end{table}

\section{Observations and data reduction\label{sec_obs}}


\subsection{\xmm\label{sec_obsxmm}}

\xmm\ observed the five INS candidates for 50 -- 80~ks each between May 2023 and April 2024 as part of programme 092128. All observations were conducted with the THIN filter. EPIC-pn \citep{2001A&A...365L..18S} was operated in full-frame (FF) mode and the two EPIC MOS detectors \citep{2001A&A...365L..27T} in small-window (SW) mode. The data reduction was conducted using the \xmm\ Science Analysis Software (SAS, version: 21.0.0). For EPIC-pn we kept single and double pattern events (PATTERN $\leq 4$), while events of all valid patterns were accepted for MOS (PATTERN $\leq 12$). We also only kept uncorrupted events by applying FLAG $=$ 0. With the exception of the observation targeting \oeon, we found that all observations were affected by periods of high background flaring. We removed these periods by applying a $3\sigma$ clipping algorithm and list the remaining clean time in Table~\ref{tab_obs}.

To improve X-ray sky localisation with respect to the initial eRASS position, we applied the \texttt{edetect\_stack} task to perform source detection using all three EPIC cameras in the five standard \xmm\ energy bands \citep{2019A&A...624A..77T,2020A&A...641A.137T}. Based on the resulting X-ray source list, we used the \texttt{eposcorr} task, with varying cuts in X-ray detection likelihood and positional accuracy, to refine the astrometry based on matches between the detected X-ray field and optical sources in the Guide Star Catalogue \citep[version: 2.4.2;][]{2008AJ....136..735L}. The refined positions in equatorial and Galactic coordinates as well as the applied shifts and rotations with respect to the initial position from the source detection are listed for all five candidates in Table~\ref{tab_pos}.

We defined the background regions to be on the same chip and have a similar RAWY value as the source region for EPIC-pn, whereas a background region on a neighbouring chip was defined for the MOS detectors. Source regions were optimised for signal-to-noise by using the \texttt{eregionanalyse} task in the 0.2 -- 12~keV band. Only for \ostt\ was a source region radius of 17" used for the extracted spectra to avoid contamination from two harder background sources. Spectra and light curves were extracted in accordance with the general SAS guidelines. We grouped the extracted spectra applying a maximum oversampling factor of 3 and including at least 25 counts per spectral bin. To enable accurate timing studies, we applied the barycentric correction with the \texttt{barycen} task using the DE405 ephemeris and the refined X-ray sky positions (Table~\ref{tab_pos}).

\begin{table*}
\caption{X-ray source detection and astrometry results\label{tab_pos}}
\centering
\scalebox{.85}{
\begin{tabular}{ccccccccccc}
\hline\hline
Source & RA & DEC & Pos.~Error\tablefootmark{(a)} & $l$\tablefootmark{(b)} & $b$\tablefootmark{(b)} & Ref.~Matches\tablefootmark{(c)} & Offset in RA & Offset in DEC & Rotation \\
& (h:m:s) & (d:m:s) & (arcsec) & (degree) & (degree) & & (arcsec) & (arcsec) & (degree) & \\
\hline
\ostt & 07:23:02.55 & -19:42:26.20 & 0.35 & 233.994970 & -2.211518 & 210 &  2.13(21) &  1.13(21) & -0.037(30)\\
\oeon & 08:19:52.32 & -13:19:31.13 & 0.31 & 235.360994 & 12.758652 &  47 &  0.53(18) & -1.15(21) & -0.021(29)\\
\oefo & 08:40:46.17 & -11:52:23.39 & 0.30 & 236.983752 & 17.823565 &  75 & -0.82(19) & -0.41(18) & -0.041(24)\\
\otfs & 13:47:25.75 & -36:34:14.98 & 0.27 & 315.304150 & 24.952107 &  62 & -1.73(16) &  0.96(14) &  0.031(20)\\
\offf & 14:45:15.96 & -37:44:28.19 & 0.35 & 326.549129 & 19.855389 &  31 & -1.28(22) & -0.56(22) & -0.104(40)\\
\hline
\end{tabular}}
\tablefoot{
\tablefoottext{a}{Radius of a circle that includes 68\% of measurements as computed from the output of \texttt{edetect\_stack} and \texttt{eposcorr} via $r_{68} = \sqrt{\frac{2.3}{2} (RADEC\_ERR^2+RAOFFERR^2+DECOFFERR^2)}$.}
\tablefoottext{b}{Galactic coordinates.}
\tablefoottext{c}{Number of potential matches between the Guide Star Catalogue \citep[version: 2.4.2;][]{2008AJ....136..735L} and EPIC source lists considered for the applied boresight correction with \texttt{eposcorr}.}
}
\end{table*}


\subsection{\eso}

The fields of the five INS candidates were observed between May and December 2023 with the \forst\ instrument at the \eso\ using the R\_SPECIAL filter (Table~\ref{tab_obs}). For each source, eight to eleven individual exposures of 400~s were obtained, adding up to total exposure times ranging from 3200~s to 4400~s. The basic data reduction was conducted with the FORS workflow for imaging data (version 5.6.2) run within the EsoReflex environment \citep{2013A&A...559A..96F}. The astrometric calibration of the reduced images was then conducted using the \texttt{astrometry.net} code \citep{2010AJ....139.1782L}. Similar to \citet{2024A&A...683A.164K, 2025A&A...694A.160K}, the saturation of many brighter field sources made it necessary to create new index files based on deeper optical imaging surveys rather than the standard \gaia\ index files available in \texttt{astrometry.net}. We consequently used data from the Legacy Survey DR10 \citep{2019AJ....157..168D} and Pan-STARRS DR1 \citep{2016arXiv161205560C} surveys for this task. Based on the improved astrometry, we used the astropy \texttt{reproject} package \citep{astropy:2013,astropy:2018,astropy:2022} to align the individual images for stacking. Lastly, we used the \texttt{ccdproc} package \citep{matt_craig_2017_1069648} to remove cosmic rays by applying the \texttt{L.A. Cosmic} algorithm \citep{2001PASP..113.1420V, curtis_mccully_2018_1482019}.

\subsection{\fast\label{sec_obsfast}}

\fast\ observed the fields of \oeon\ and \oefo\ in September and October 2024 for 1200\,s and 2700\,s, respectively.
Observations were carried out with the central beam of \fast\ 19-beam receiver covering a frequency range from 1.0 to 1.5\,GHz, channelised into 1024 channels with a 0.488\,MHz resolution \citep{2019SCPMA..6259502J}.
The data were taken in \texttt{Tracking} mode with a $49.152$~µs sampling time and stored in search-mode PSRFITS format \citep{2004PASA...21..302H}.

To search over a wide range of possible radio pulsations, we employed the pulsar search software package \textsc{PRESTO}\footnote{\url{https://github.com/scottransom/presto}} \citep{2002AJ....124.1788R}, following its standard procedures for both periodicity and single-pulse searches. Radio frequency interference (RFI) was first identified and masked using the \texttt{rfifind} routine. The applied data length in the RFI search was 2.0\,s. The data were then de-dispersed over a range of trial dispersion measures (DMs) using \texttt{prepsubband}. We searched for radio pulsations over a DM range of 0--1400\,pc\,cm\(^{-3}\), significantly exceeding the maximum values (\(\sim100\)\,pc\,cm\(^{-3}\)) predicted by the Galactic electron density models NE2001 \citep{2002astro.ph..7156C} and YMW16 \citep{2017ApJ...835...29Y} along the lines of sight to the two target sources. The DM step sizes were determined using \texttt{DDplan.py}, with values of 0.05, 0.10, 0.20, 0.30, and 0.50\,pc\,cm\(^{-3}\) adopted over the DM ranges of 0--113.7, 113.7--189.5, 189.5--341.1, 341.1--586.5, and 586.5--1404.5\,pc\,cm\(^{-3}\), respectively. The resulting de-dispersed time series were transformed into the frequency domain via \texttt{realfft}. The effects of low-frequency noise in the power spectra were removed using \texttt{rednoise}. We performed searches for possible periodic signals using \texttt{accelsearch}, which applies Fourier-domain acceleration algorithms in the Fourier spectra. The maximum absolute value of acceleration (defined by the $z_{\rm max}$ option in \texttt{accelsearch}) was set to 20 and 200, which represents the number of Fourier frequency bins that a signal can drift during an individual observation. All identified periodic signals from all DM trials were shifted using a customised version of \texttt{ACCEL\_sift.py}. Promising candidates were then folded using the \texttt{prepfold} routine to produce diagnostic plots for further visual inspection. In parallel, the \texttt{single\_pulse\_search.py} was used to search for single pulses in all de-dispersed time series, adopting a signal-to-noise (S/N) threshold of 8. Any possible single radio pulses were displayed using \texttt{waterfaller.py} and were visually inspected.


\section{Results\label{sec_analysis}}

\begin{figure*}
\begin{minipage}[t]{0.33\textwidth}
\vspace*{-5.97cm}\includegraphics[width=\linewidth]{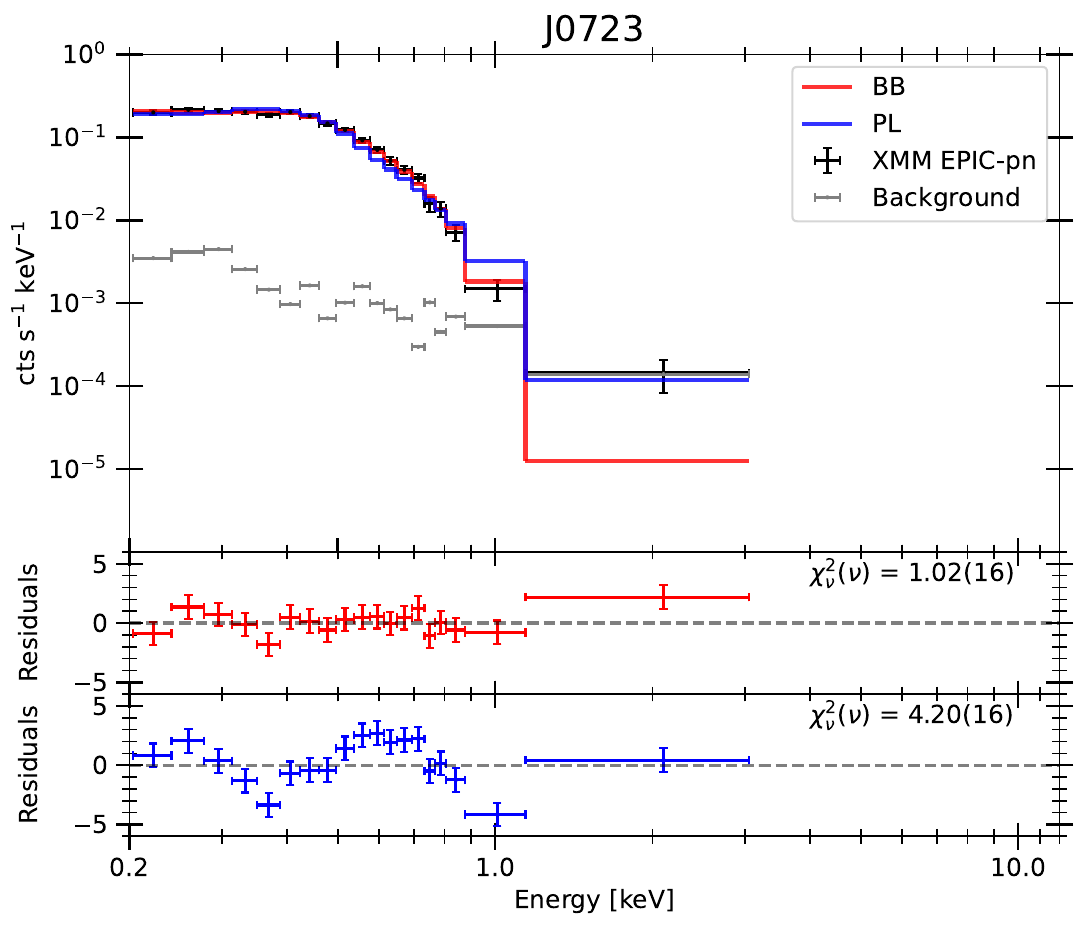}
\end{minipage}
\begin{minipage}[t]{0.33\textwidth}
\vspace*{-5.97cm}\includegraphics[width=\linewidth]{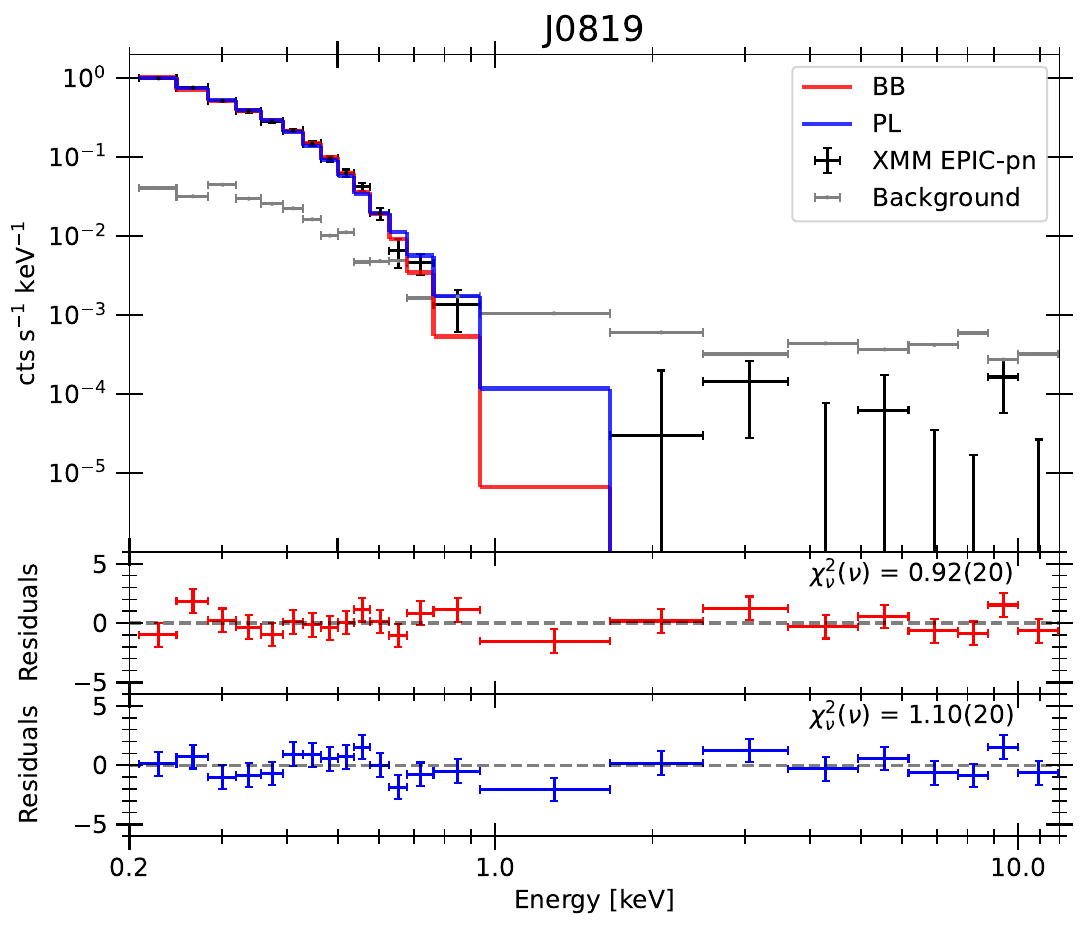}
\end{minipage}
\begin{minipage}[t]{0.33\textwidth}
\includegraphics[width=\linewidth]{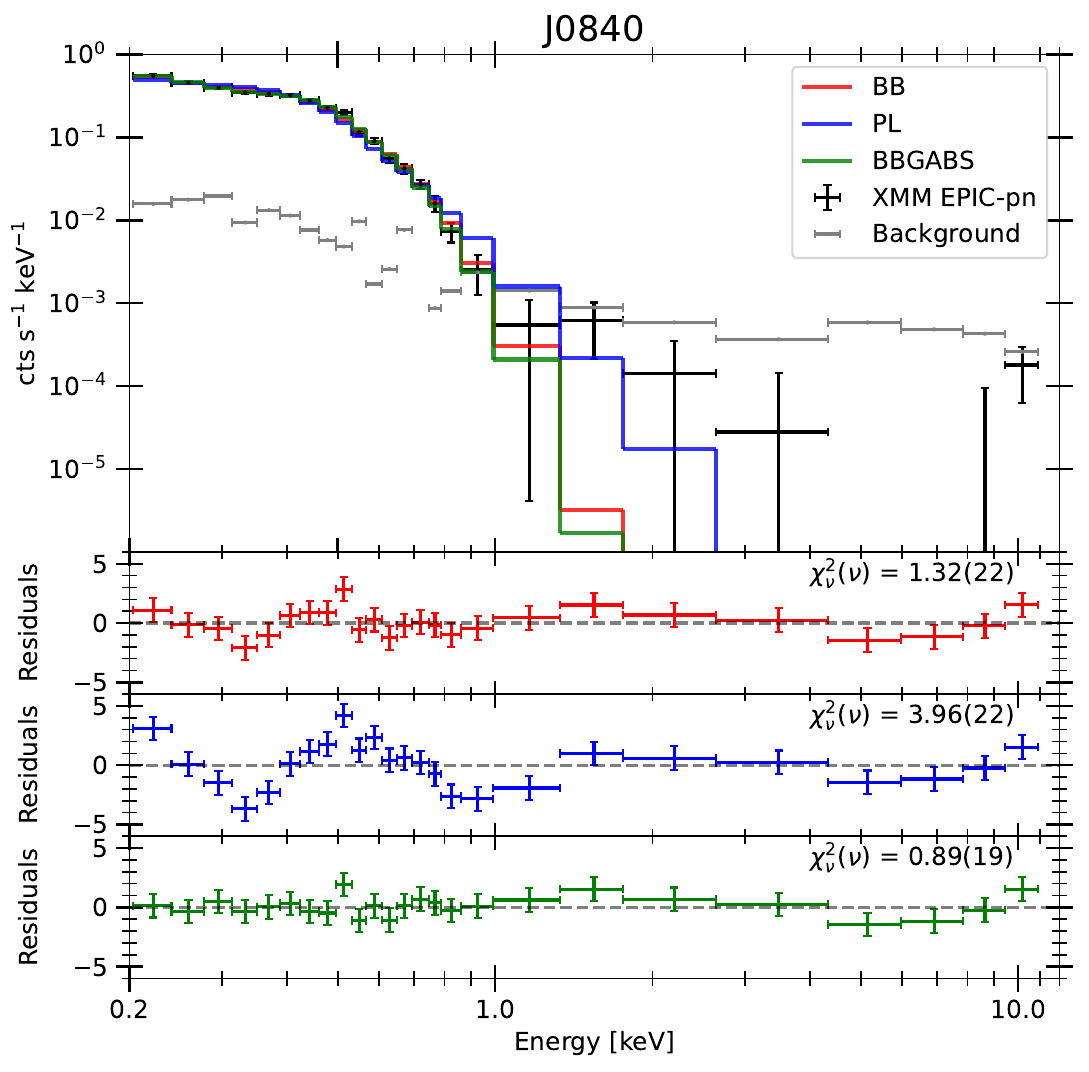}
\end{minipage}

\begin{minipage}[t]{0.33\textwidth}
\includegraphics[width=\linewidth]{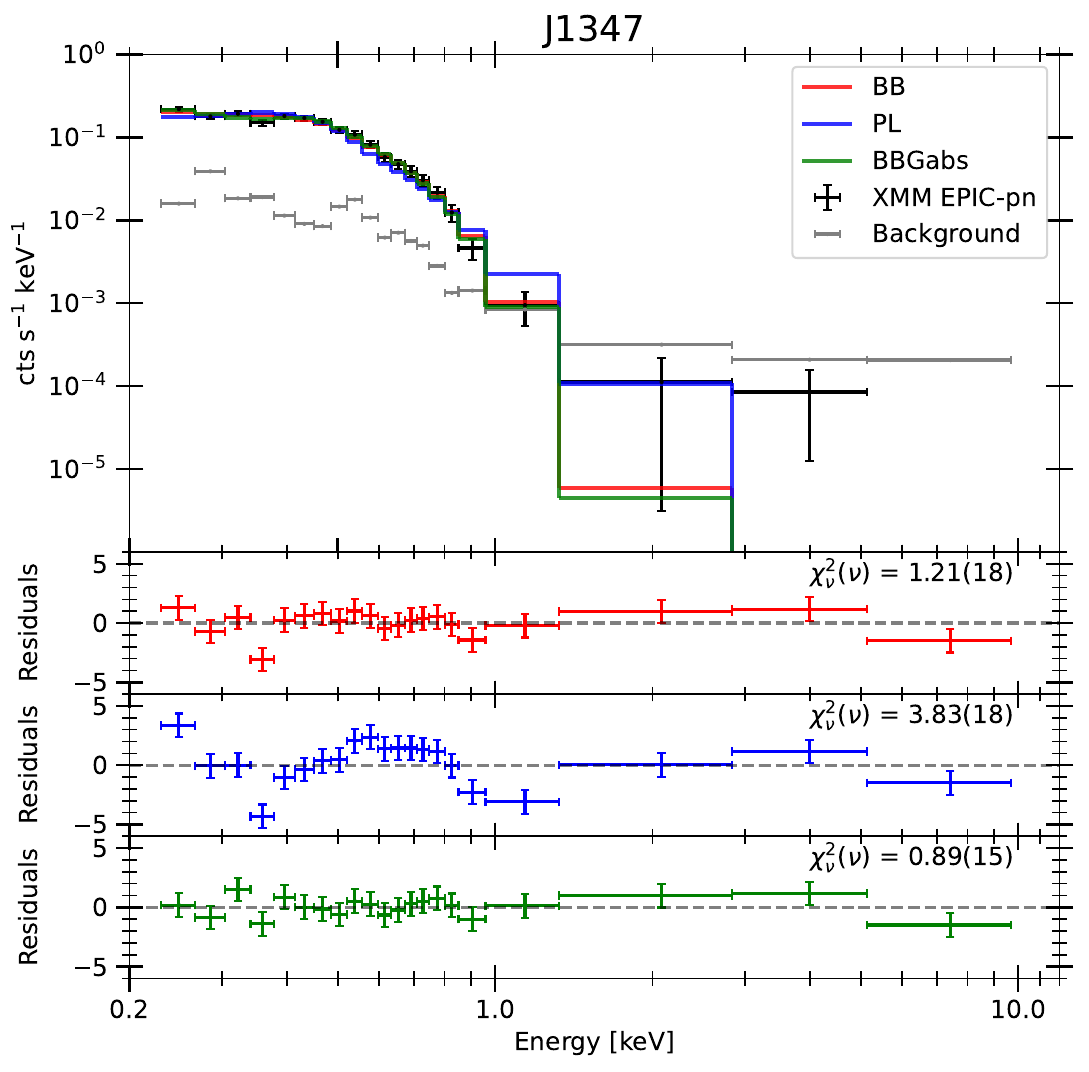}
\end{minipage}
\begin{minipage}[t]{0.33\textwidth}
\vspace*{-5.97cm}\includegraphics[width=\linewidth]{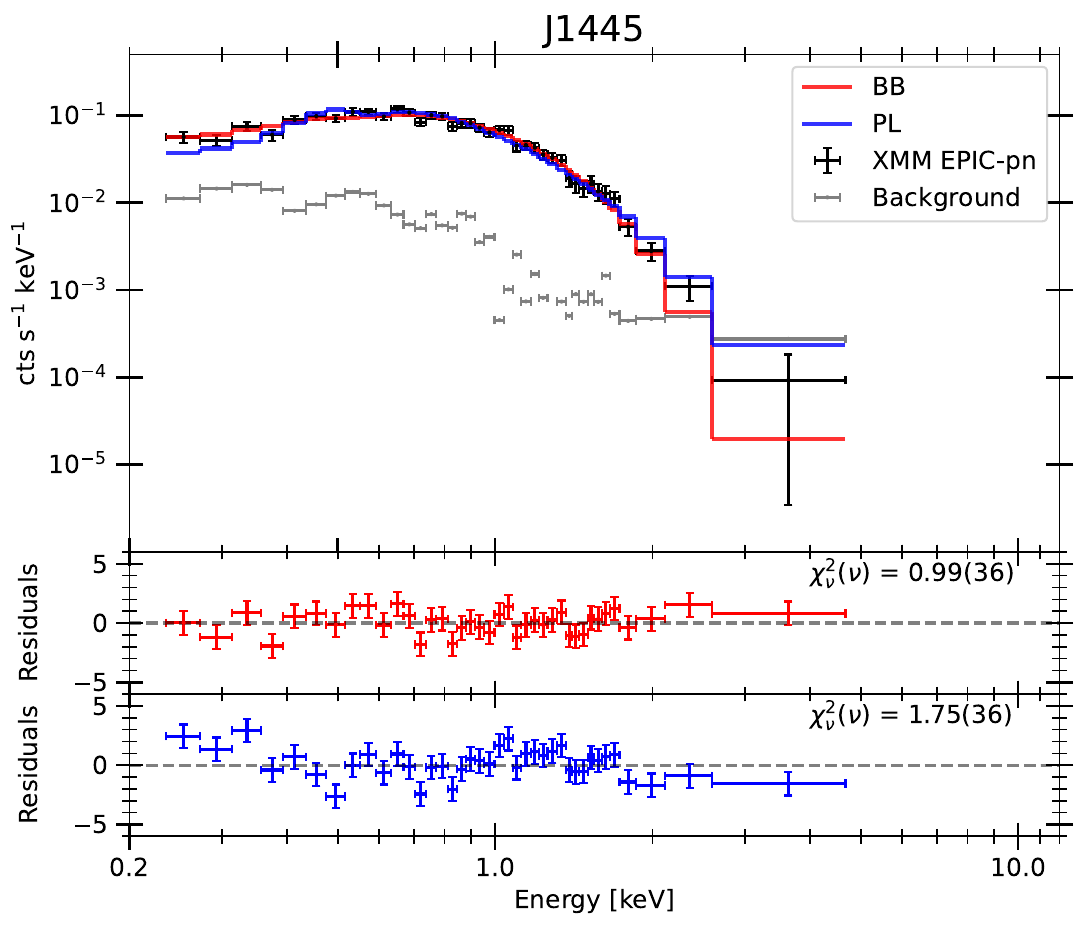}
\end{minipage}

\caption{EPIC-pn spectra and best-fit spectral models of all five candidates. The spectral parameters of the presented fits are listed in Table~\ref{tab_fitres}. }
\label{fig_spec_fits}
\end{figure*}



\subsection{X-ray spectral analysis\label{sec_x_ray_fits}}

We studied the spectra of the five INS candidates by modelling them with the X-ray spectral fitting tool XSPEC \citep[Version: 12.14.1d;][]{1996ASPC..101...17A}. We applied the chi-square statistic in the model optimisation. To account for interstellar absorption, we used the elemental abundances from \citet{2000ApJ...542..914W}, and all spectral models were multiplied with a \texttt{tbabs} component. If data from multiple instruments (EPIC-pn and MOS1/2) were simultaneously fitted, the spectral model was expanded by a constant factor to account for remaining cross-calibration uncertainties. The obtained spectral parameters and EPIC-pn spectra are presented in Table~\ref{tab_fitres} and Fig.~\ref{fig_spec_fits}, respectively.

\begin{table*}
\caption{X-ray spectral analysis results
\label{tab_fitres}}
\centering
\scalebox{.77}{
\begin{tabular}{cccccccccccccc}
\hline\hline
\multicolumn{5}{l}{\ostt}\\
\hline
Instruments & Model & $\nh$ & $N_\mathrm{H, gal}$\tablefootmark{(a)} & Dist.\tablefootmark{(b)} & $kT$ & $R$\tablefootmark{(c)} & $\Gamma$\tablefootmark{(d)} & $\epsilon$\tablefootmark{(e)} & $\sigma$ & $EW$\tablefootmark{(f)} & $\chi^2_\nu(\nu)$ & Absorbed flux\tablefootmark{(g)}\\
 &  &  $[10^{20}$\,cm$^{-2} ] $ &  $[10^{20}$\,cm$^{-2} ] $ & [kpc] & [eV] & [km] & & [eV] & [eV] & [eV] & & [$10^{-13}$\,\fluxcgs]\\
\hline\noalign{\smallskip}
EPIC-pn & \texttt{BB} & $4.9^{+0.7}_{-0.6}$  & 64.7 & $<1.4$ & $71.1^{+1.6}_{-1.6}$ & $4.5^{+0.6}_{-0.5}$ & & & & &  1.02(16) & $1.292^{+0.023}_{-0.023}$\\
EPIC-pn & \texttt{PL} & $16.3^{+1.1}_{-1.0}$ & 64.7 & $<2.1$ &  & &  $7.75^{+0.2}_{-0.19}$ & & & & 4.20(16) & $1.220^{+0.022}_{-0.022}$\\
All & \texttt{BB} & $4.6^{+0.6}_{-0.5}$ & 64.7 & $<1.4$ & $71.6^{+1.4}_{-1.4}$ & $4.3^{+0.5}_{-0.4}$ & & & & & 1.04(41) & $1.320^{+0.020}_{-0.020}$ \\

\hline
\multicolumn{5}{l}{\oeon}\\
\hline
Instruments & Model & $\nh$ & $N_\mathrm{H, gal}$\tablefootmark{(a)} & Dist.\tablefootmark{(b)} & $kT$ & $R$\tablefootmark{(c)} & $\Gamma$\tablefootmark{(d)} & $\epsilon$\tablefootmark{(e)} & $\sigma$ & $EW$\tablefootmark{(f)} & $\chi^2_\nu(\nu)$ & Absorbed flux\tablefootmark{(g)}\\
 & & $[10^{20}$\,cm$^{-2} ] $ &  $[10^{20}$\,cm$^{-2} ] $ & [kpc] & [eV] & [km] & & [eV] & [eV] & [eV] & & [$10^{-13}$\,\fluxcgs]\\
 \hline\noalign{\smallskip}
EPIC-pn & \texttt{BB} & $0.6^{+0.4}_{-0.4}$    & 6.7  & $<0.3$ & $47.6^{+1.3}_{-1.3}$ & $15.8^{+2.9}_{-1.9}$ & & & & & 0.92(20) & $3.97^{+0.06}_{-0.06}$ \\
EPIC-pn & \texttt{PL} & $7.2^{+0.6}_{-0.6}$    & 6.7  &        & & & $9.21^{+0.27}_{-0.25}$ & & & & 1.10(20) & $3.58^{+0.05}_{-0.05}$ \\
All & \texttt{BB} & $0.54^{+0.4}_{-0.29}$ & 6.7 & $<0.3$ & $48.0^{+1.1}_{-1.1}$ & $15.2^{+2.4}_{-1.7}$ & & & & & 0.97(45) & $3.88^{+0.05}_{-0.05}$\\
\hline
\multicolumn{5}{l}{\oefo}\\
\hline
Instruments & Model & $\nh$ & $N_\mathrm{H, gal}$\tablefootmark{(a)} & Dist.\tablefootmark{(b)} & $kT$ & $R$\tablefootmark{(c)} & $\Gamma$\tablefootmark{(d)} & $\epsilon$\tablefootmark{(e)} & $\sigma$ & $EW$\tablefootmark{(f)} & $\chi^2_\nu(\nu)$ & Absorbed flux\tablefootmark{(g)}\\
 & &  $[10^{20}$\,cm$^{-2} ] $ &  $[10^{20}$\,cm$^{-2} ] $ & [kpc] & [eV] & [km] & & [eV] & [eV] & [eV] & & [$10^{-13}$\,\fluxcgs]\\
\hline\noalign{\smallskip}
EPIC-pn & \texttt{BB} & $2.0^{+0.5}_{-0.4}$  & 5.71 & $<1.3$ & $67.2^{+1.4}_{-1.4}$ & $5.4^{+0.6}_{-0.5}$ & & & & & 1.32(22) & $2.27^{+0.04}_{-0.04}$ \\
EPIC-pn & \texttt{PL} & $10.5^{+0.7}_{-0.6}$ & 5.71 &     & & & $7.44^{+0.18}_{-0.17}$ & & & & 3.96(22) & $2.1^{+0.04}_{-0.04}$ \\
EPIC-pn & \texttt{BBGABS} & $2.4^{+0.6}_{-1.2}$  & 5.71 & $<1.3^{\star}$ & $63.5^{+2.2}_{-2.3}$ & $7.1^{+1.6}_{-1.0}$ & & $359^{+17}_{-50}$ & $29^{+50}_{-28}$ & $30^{+60}_{-22}$ & 0.89(19) & $2.32^{+0.04}_{-0.04}$ \\
All & \texttt{BBGABS} & $<1.1$ & 5.71 & $<1.2$ & $67.0^{+1.6}_{-1.8}$ & $5.2^{+0.9}_{-0.5}$ & & $282^{+29}_{-22}$ & $67^{+14}_{-22}$ & $90^{+40}_{-50}$ &  1.15(49) & $2.50^{+0.04}_{-0.04}$ \\
\hline
\multicolumn{5}{l}{\otfs}\\
\hline
Instruments & Model & $\nh$ & $N_\mathrm{H, gal}$\tablefootmark{(a)} & Dist.\tablefootmark{(b)} & $kT$ & $R$\tablefootmark{(c)} & $\Gamma$\tablefootmark{(d)} & $\epsilon$\tablefootmark{(e)} & $\sigma$ & $EW$\tablefootmark{(f)} & $\chi^2_\nu(\nu)$ & Absorbed flux\tablefootmark{(g)}\\
 & & $[10^{20}$\,cm$^{-2} ] $ &  $[10^{20}$\,cm$^{-2} ] $ & [kpc] & [eV] & [km] & & [eV] & [eV] & [eV] & & [$10^{-13}$\,\fluxcgs]\\
\hline\noalign{\smallskip}
EPIC-pn & \texttt{BB} & $2.6^{+0.9}_{-0.8}$  & 4.32 & $<1.3^{\star}$ & $80.4^{+2.3}_{-2.3}$ & $2.51^{+0.4}_{-0.28}$ & & & & & 1.21(18) & $1.192^{+0.026}_{-0.026}$ \\
EPIC-pn & \texttt{PL} & $14.6^{+1.5}_{-1.4}$ & 4.32 &        & & & $7.05^{+0.25}_{-0.23}$ & & & & 3.83(18) & $1.051^{+0.024}_{-0.024}$ \\
EPIC-pn & \texttt{BBGABS} & $2.2^{+0.9}_{-0.8}$  & 4.32 & $<1.7$ & $78.0^{+2.5}_{-2.5}$ & $2.8^{+0.5}_{-0.4}$ & & $373^{+15}_{-15}$ & $3.1^{+40}_{-1.0}$ & $27^{+400}_{-14}$ & 0.89(15) & $1.263^{+0.028}_{-0.028}$ \\
All & \texttt{BBGABS} & $1.45^{+0.28}_{-0.6}$ & 4.32 & $<1.1$ & $79.6^{+1.9}_{-1.9}$ & $2.53^{+0.3}_{-0.24}$ & & $363^{+12}_{-12}$ & $2.1^{+40}_{-0.7}$ & $23^{+300}_{-17}$ & 1.08(50) & $1.357^{+0.025}_{-0.025}$ \\
\hline
\multicolumn{5}{l}{\offf}\\
\hline
Instruments & Model & $\nh$ & $N_\mathrm{H, gal}$\tablefootmark{(a)} & Dist.\tablefootmark{(b)} & $kT$ & $R$\tablefootmark{(c)} & $\Gamma$\tablefootmark{(d)} & $\epsilon$\tablefootmark{(e)} & $\sigma$ & $EW$\tablefootmark{(f)} & $\chi^2_\nu(\nu)$ & Absorbed flux\tablefootmark{(g)}\\
 &  & $[10^{20}$\,cm$^{-2} ] $ &  $[10^{20}$\,cm$^{-2} ] $ & [kpc] & [eV] & [km] & & [eV] & [eV] & [eV] & & [$10^{-13}$\,\fluxcgs]\\
\hline\noalign{\smallskip}
EPIC-pn & \texttt{BB} & $3.2^{+1.1}_{-1.0}$  & 5.77 & $<1.1^{\star}$ & $210^{+5}_{-5}$ & $0.281^{+0.022}_{-0.018}$ & & & & & 0.99(36) & $1.332^{+0.027}_{-0.027}$ \\
EPIC-pn & \texttt{PL} & $42.0^{+2.9}_{-2.8}$ & 5.77 & & & & $4.88^{+0.17}_{-0.16}$ & & & & 1.75(36) & $1.322^{+0.027}_{-0.027}$ \\
All & \texttt{BB} & $2.7^{+0.9}_{-0.9}$ & 5.77 & $<1.2$ & $206^{+4}_{-4}$ & $0.288^{+0.018}_{-0.016}$ & & & & & 1.26(94) & $1.327^{+0.022}_{-0.022}$\\
\noalign{\smallskip}\hline
\end{tabular}}
\tablefoot{$1\sigma$ confidence intervals are provided for the best-fit parameters.
\tablefoottext{a}{Galactic column density in the direction of the sources as inferred from \citet{2016A&A...594A.116H}.}
\tablefoottext{b}{Upper distance limit based on the lower confidence level limit of the $N_\mathrm{H,E(B-V)}$ value in the direction of the source equal to the 99\% confidence region upper limit of the best-fit $\nh$ value. Distance values were inferred from the online tool described in \citet{2024arXiv240303127D}. Values marked with a star ($\star$) indicate cases where the upper limit on $\nh$ is always above the lower confidence limit on $N_\mathrm{H,E(B-V)}$ regardless of distance, indicating a possible extragalactic nature. Here, we give the distance according to the absolute $N_\mathrm{H,E(B-V)}$ value.}
\tablefoottext{c}{A distance of 1\,kpc assumed for the \texttt{BB} emission radius at infinity.}
\tablefoottext{d}{Photon index as given by the XSPEC \texttt{powerlaw} model component.}
\tablefoottext{e}{Central line energy of the Gaussian absorption component.}
\tablefoottext{f}{The equivalent width estimated from $\int \frac{f_c-f_o}{f_c} dE$, with $f_c$ being the continuum and $f_o$ the observed flux. The errors provide the maximum and minimum EW values obtained from all possible combinations of the upper and lower $1\sigma$ confidence interval limits of the model parameters.}
\tablefoottext{g}{Absorbed model flux in the $0.2-12$\,keV range.}
}
\end{table*}

We began the spectral analysis by applying a simple absorbed blackbody (\texttt{BB}) model to the EPIC-pn spectra of all sources. For \ostt, \oeon, and \offf, we find this model to fit the spectra well as indicated by the good fit statistics close to one ($\chi^2_\nu (\nu) \sim 1.02 (16)$, $\chi^2_\nu (\nu) \sim 0.92 (20)$, and $\chi^2_\nu (\nu) \sim 0.99 (36)$; see Table~\ref{tab_fitres}). For the two remaining sources (\oefo\ and \otfs), the \texttt{BB} fits indicate remaining residuals mostly at low energies ($<400$~eV, Fig.~\ref{fig_spec_fits}). They can be best accounted for by including a narrow ($\sigma\sim 2-80$~eV) Gaussian absorption component (\texttt{GABS}) in the \texttt{BB} spectral model at $300-400$~eV. While for \oefo\ this component allows the removal of the weak structured residuals ($1-2\sigma$ deviation) observed in the \texttt{BB} fit, in the case of \otfs\ the line appears to model only the residual caused by the $3-4\sigma$ deviation of a single spectral bin (the fourth from the left in the \texttt{BB} fit residuals presented in Fig.~\ref{fig_spec_fits}).

While lines of similar width and strength, as in \oefo\ and \otfs, have been observed in other thermally emitting INSs \citep[e.g.][]{2017MNRAS.468.2975B}, it is important to explore their significance and whether they may arise due to fluctuations introduced by the counting statistics. To this end, we conducted simulations based on the best-fit results in Table~\ref{tab_fitres} to estimate the false-positive (fit indicates a feature that is not truly contained in the data) and false-negative (a contained feature is not identified by the spectral fit) rates. From the simulation of a thousand \texttt{BB} spectra (based on the EPIC-pn best-fit solutions listed in Table~\ref{tab_fitres}), we find in 0.7\% (\oefo) and 2.6\% (\otfs) of all cases that the inclusion of a Gaussian absorption line leads to an improvement of similar size or larger in the fit statistic as observed here. These low false-positive rates imply that it is unlikely that the observed features arise solely due to counting statistics. From the simulation of a thousand \texttt{BBGABS} spectra (with the same parameters as the EPIC-pn best-fit solutions in Table~\ref{tab_fitres}) we infer that the lines are not recovered in 15.5\% and 22.8\% of all simulations. While these false-negative rates indicate that additional observations under similar conditions may miss the absorption lines, the probabilities are low enough to imply that the conducted observations overall do allow for their detection. We conclude that spectral variations due to counting statistics alone are unlikely to cause the observed features.

Alternatively, the features could also originate from an improper estimation and subtraction of the background. To test the background's influence, we varied the background region in size and location and found this to alter the values of the best-fit reduced chi-square of the \texttt{BB} and \texttt{BBGABS} models, even though the resulting fit parameters are always consistent (within $1-2\sigma$) with those listed in Table~\ref{tab_fitres}. Regardless of the chosen background region, we find that the addition of the \texttt{GABS} component always significantly improves the spectral fit for both sources. Consequently, the lines must be intrinsic to the two sources or originate from background variations in the source region, which are difficult to probe.

We found that non-thermal components did not allow us to convincingly model the observed spectral emission. This is exemplarily presented by the simplest model: a single absorbed power-law (\texttt{PL}). As shown in Fig.~\ref{fig_spec_fits} and Table~\ref{tab_fitres}, for all sources a \texttt{PL} continuum leaves significant residuals and converges to unusually steep spectral slopes ($\Gamma \gtrsim 5$). Consequently, the bulk of the emission for all INS candidates appears to be of thermal nature.

The $\nh$ values of all the thermal fits are generally in agreement with a Galactic nature as they are comparable to or below the total Galactic column density in the direction of the sources based on the 2D HI4PI maps \citep[Table~\ref{tab_fitres};][]{2016A&A...594A.116H}. Using the 3D-$\nh$ tool \citep{2024arXiv240303127D}, we estimated upper distance limits based on the 99\% confidence intervals of the best-fit $\nh$ values, yielding distances ranging from 1.1 to 1.7~kpc (Table~\ref{tab_fitres}). Only for \oeon, we infer a distance $<300$~pc.

For most of the sources, the obtained temperature values ($\sim45 - 80$~eV) are in agreement with those observed for other thermally emitting INSs \citep{2020MNRAS.496.5052P}. Only for \offf\ the best-fit \texttt{BB} temperature of $\sim 210$~eV and radius of $\sim 280$~m at a 1~kpc distance may imply that the observed X-ray emission originates from a small heated region. Within the distance limits inferred from interstellar medium (ISM) absorption, the emission region sizes of the other sources are also below the canonical INS radius of $\sim12$~km. These results resemble the radius estimates from absorbed \texttt{BB} fits to the spectra of the predominantly thermally emitting XDINSs \citep[e.g.][]{2022MNRAS.516.4932D}. For many thermally emitting INSs, multiple \texttt{BB} components need to be combined to properly fit the thermal continuum \citep[e.g.][]{2019PASJ...71...17Y,2022A&A...661A..41S}. Given that the single \texttt{BB} models fit the continuum well for all sources, we found that models combining multiple \texttt{BB} components are not required to further improve the spectral fitting. We conclude that higher signal-to-noise data is needed to resolve hot polar caps or the non-uniform surface temperature distribution for these sources.

Next to \texttt{BB} fits, neutron star atmosphere models \citep[\texttt{NSA};][]{1996A&A...315..141Z,1995ASIC..450...71P} also allow us to fit the thermal continuum of all five candidates well ($\chi^2_\nu \sim 1$). Assuming canonical neutron stars with a radius of 12~km and a mass of $1.4$~M$_\odot$, we found that non-magnetised models appear to be favoured and that the fits converge to overall smaller temperature values ($kT\sim 10-40$~eV for most sources, $kT\sim 90-150$~eV for \offf). The obtained $\nh$ values are higher for \ostt\ ($7.6^{+0.6}_{-0.5}\times 10^{20}$~cm$^{-2}$) and \offf\ ($10.19^{+1.6}_{-1.6}\times 10^{20}$~cm$^{-2}$) but are in agreement with the \texttt{BB} results for the remaining sources. At the same time, the \texttt{NSA} models converge to small distance estimates ($\lesssim 40-300$~pc) for \ostt, \oeon, and \oefo, whereas greater distances can be obtained in magnetised \texttt{NSA} model fits to the spectra of the remaining sources. Similar to a \texttt{BB} continuum, the \texttt{NSA} model also requires absorption line components in the fits to \oefo\ and \otfs.

Finally, we also tried fits including the spectra from MOS1 and MOS2 (Table~\ref{tab_fitres}). We find including the MOS data results in an overall broader spread of the residuals around the best-fit solution, leading to larger values of the fit statistic for most sources. The multiplicative constant that was included to model the remaining deviations between the single instruments was not able to fully account for this spread. The inclusion of additional model components beyond the best-fit solutions shown in Table~\ref{tab_fitres} does not allow further improvement of the spectral modelling. Despite the larger fit statistic, the resulting parameter values from the MOS instruments are fully consistent with the EPIC-pn only results (Table~\ref{tab_fitres}).


\subsection{Limits on high-energy excess\label{sec_x_ray_excess}}

\begin{table*}
\small
\caption{Upper limits on hard X-ray excess
\label{tab_excess}}
\centering
\scalebox{1.}{
\begin{tabular}{l|ccc|ccc}
\hline\hline\noalign{\smallskip}
 & \multicolumn{3}{c}{Single source PSF fitting\tablefootmark{(a)}} & \multicolumn{3}{c}{Multiple source PSF fitting\tablefootmark{(b)}}\\
Source & DET\_ML & Flux$_{1-12\mathrm{keV}}$ &$F_{1-12\mathrm{keV}}/F_{BB}$ & DET\_ML & Flux$_{1-12\mathrm{keV}}$ & $F_{1-12\mathrm{keV}}/F_{BB}$\\
 & & [$10^{-15}$\,\fluxcgs] & [\%] & &  [$10^{-15}$\,\fluxcgs] & [\%] \\
\hline
\ostt & $1.99$    & $2.0 \pm 1.0$ & $0.55^{+0.28}_{-0.28}$ & $0$        & $0.0 \pm 0.5$ & $0.0^{+0.14}_{-0.14}$ \\
\oeon & $0.07$    & $0.7 \pm 0.8$ & $0.14^{+0.16}_{-0.16}$ & $0.07$     & $0.7 \pm 0.8$ & $0.14^{+0.16}_{-0.16}$ \\
\oefo & $2.62$    & $2.2 \pm 1.0$ & $0.44^{+0.21}_{-0.21}$ & $2.605$    & $2.1 \pm 1.0$ & $0.42^{+0.21}_{-0.21}$ \\
\otfs & $0.60$    & $1.3 \pm 0.9$ & $0.6^{+0.5}_{-0.5}$    & $0.537$    & $1.2 \pm 0.9$ & $0.5^{+0.5}_{-0.5}$ \\
\offf & $1516.55$ & $117 \pm 5$   & $74^{+4}_{-4}$         & $1540.319$ & $117 \pm 5$   & $74^{+4}_{-4}$ \\
\hline
\end{tabular}}
\tablefoot{Only photons between 1~keV and 12~keV are considered. Errors give $1\sigma$ confidence levels.
\tablefoottext{a}{PSF fitting conducted at the position of the target, not accounting for any nearby sources.}
\tablefoottext{b}{All nearby field sources (within 1\arcmin) included in the PSF fitting at the position of the candidates. Field sources are identified based on the results from a previous \texttt{edetect\_stack} run.}
}
\end{table*}

For all candidates, the spectral analysis implies that the bulk of the observable X-ray emission originates from thermal components. Nevertheless, it is important to quantify the current limits on the existence of non-thermal emission components that may dominate the INS emission at higher energies. In order to achieve reliable limits on high-energy excess emission, we first applied the \texttt{edetect\_stack} task to do a source detection in the 1 -- 12~keV band where we expect the soft thermal emission components to be faint. We then performed point spread function (PSF) fitting in the same energy band on the EPIC-pn data using the \texttt{emldetect} task, assuming either a single input source at the candidate's position, or inserting a source at that position while also including all nearby sources (within 1\arcmin) detected in the previous \texttt{edetect\_stack} run. We note that only \offf\ was recovered in the source detection above 1~keV; therefore, no artificial source was injected for this candidate in the second case. To convert the resulting count rates into flux values, we computed energy conversion factors (ECF) by extracting spectra at the position of each candidate, assuming a circular extraction region with a radius of $1$\arcmin. We then applied the corresponding response files and XSPEC to estimate the ECF in the 1 -- 12~keV band based on a power-law spectrum with the same $\nh$ values as the best-fit solution of each source in Table~\ref{tab_fitres} and a photon index of 2. We present the resulting detection likelihood (DET\_ML) and flux values in Table~\ref{tab_excess}.

The only candidate to be significantly detected above $1$~keV is \offf, as the DET\_ML values for all other candidates are quite low ($<3$). Consequently, their observed flux values are within 1 -- 2$\sigma$ in agreement with zero. We computed the ratio of the obtained values for the remaining flux at higher energies to the best-fit unabsorbed \texttt{BB} model flux in order to study the relative strength between the observed thermal emission and possible non-thermal emission components in these sources. For all four targets undetected above 1~keV, the resulting ratios are below 1\% (Table~\ref{tab_excess}). While such ratios may appear low, a similarly low value of $\sim 0.3\%$ was for example observed for the RPP \tmzsfs\ , which has a well detectable high-energy component \citep{2005ApJ...623.1051D}. For the two XDINSs \magzf\ and \magoe\ non-thermal spectral components were also recently discovered \citep{2020ApJ...904...42D, 2022MNRAS.516.4932D}. Using the best-fit values from \citet{2022MNRAS.516.4932D}, we computed $F_{1-12\mathrm{keV}}/F_{BB}$ fractions of 0.87\% and 0.03\% for \magzf\ and \magoe, respectively. Comparing the values in Table~\ref{tab_excess}, the current observations allow us to detect a non-thermal component, as observed in \magzf\ for \ostt\ and \oeon, but are overall still too shallow to discover a weak non-thermal component, as observed in \magoe. Consequently, deeper X-ray observations will be needed to fully exclude the existence of non-thermal emission components for the five sources discussed here.

For \offf\ the higher temperature \texttt{BB} is significantly detectable up to $\sim3$~keV. Repeating the source detection in the 3 -- 12~keV band, we found \offf\ to be undetected (DET\_ML $=0$) and consequently estimated a flux upper limit of $(0.8\pm2.4)\times 10^{-15}$~\fluxcgs\ (3 -- 12~keV). This implies a ratio of $(0.5 \pm 1.6)$\% between the flux contained in an undetected high-energy component (3 -- 12~keV) and in the observed \texttt{BB}-like spectrum.


\subsection{X-ray short- and long-term variability \label{sec_x_ray_time}}

\begin{figure}[t]
\includegraphics[width=\linewidth]{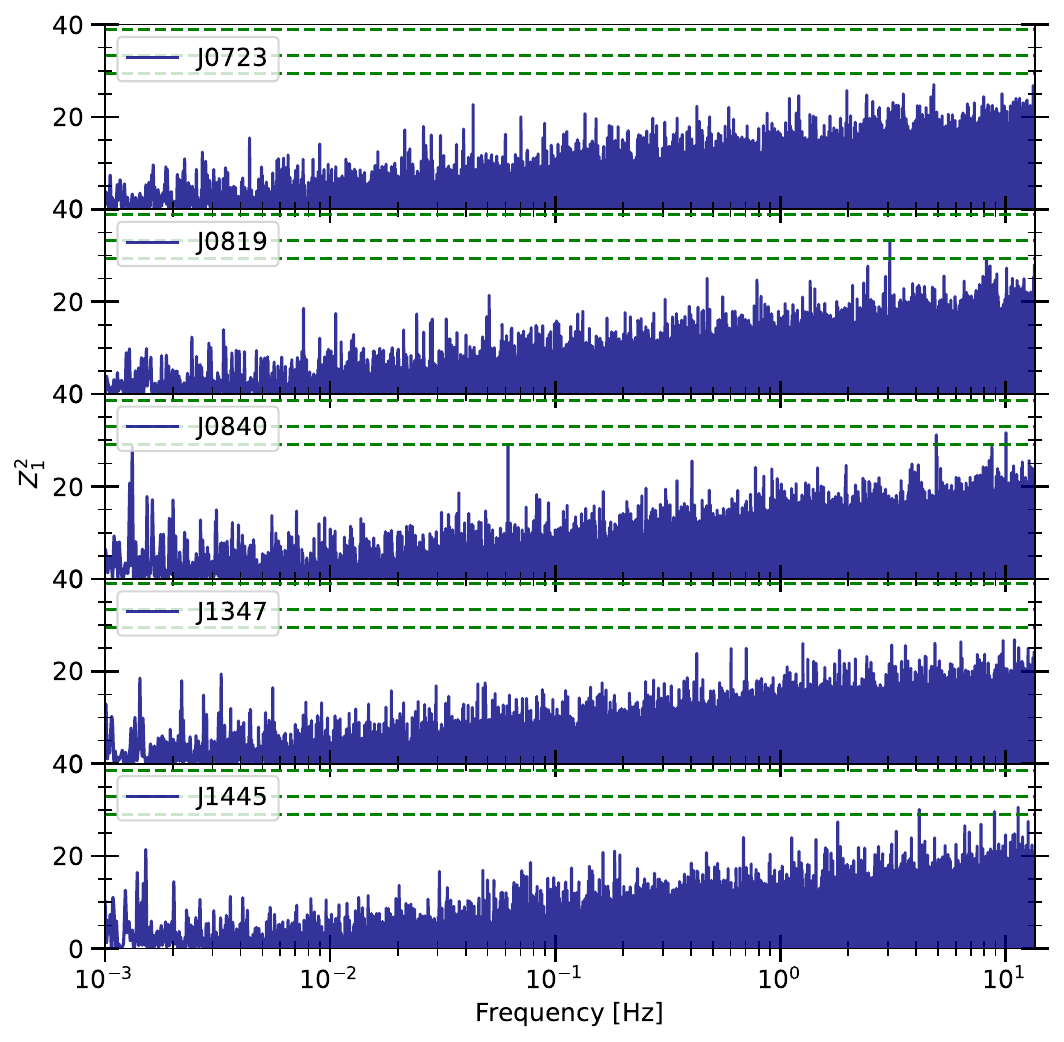}\\
\caption{Power spectra from a $Z^2_1$-search in the $10^{-3}-13.5$~Hz band using the event arrival times from EPIC-pn. The dashed green lines indicate the $1-3\sigma$ significance levels (from lowest to highest). }
\label{fig_powspec_fits}
\end{figure}


\begin{figure}[t]
\includegraphics[width=\linewidth]{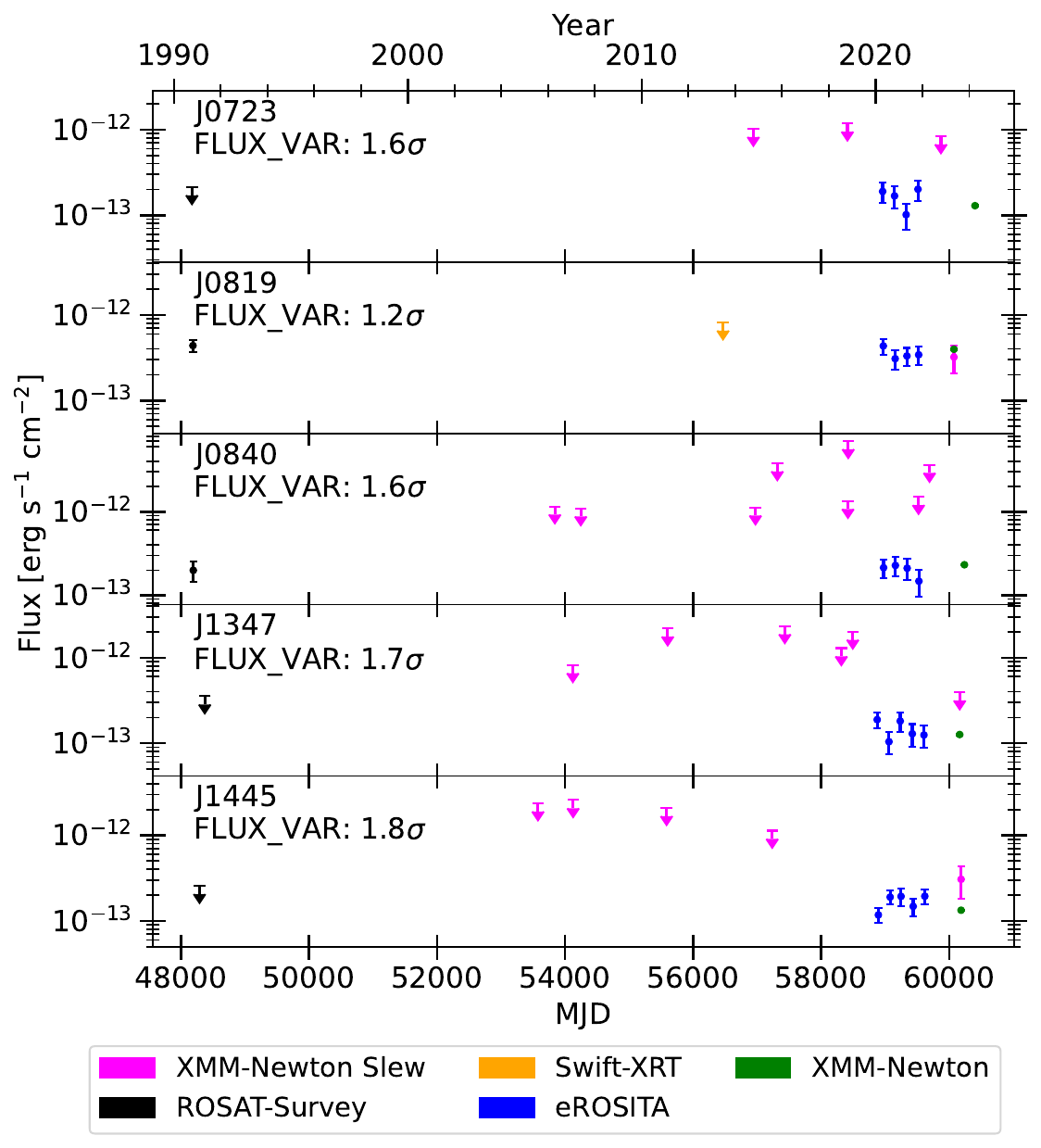}\\
\caption{Long-term X-ray light curves in the 0.2 -- 12~keV band, including flux values and $3\sigma$ upper limits from archival X-ray observations and measurements obtained from spectral fitting (green; Table~\ref{tab_fitres}).}
\label{fig_long_term_lc}
\end{figure}


We applied the $Z^2_1$-test \citep[][]{1983A&A...128..245B} to search the EPIC-pn observations of the five candidates for short-term periodic modulations from $10^{-3}$ to $13.5$~Hz. To this end, we extracted good time interval (GTI) filtered event lists in the 0.2 -- 1~keV band. Only for \offf, which possesses the hardest spectrum of all the candidates, we considered events in the 0.2 -- 5~keV range.

For all candidates, the resulting power spectra contain no significant timing signals (see Fig.~\ref{fig_powspec_fits}). Further searches, including higher harmonics in the $Z_m$-test, using the Lomb-Scargle \citep{1976Ap&SS..39..447L,1982ApJ...263..835S} or the Bayesian-based folding method described in \citet[][]{1996ApJ...473.1059G}, also failed to discover any significant modulations in the data. Similarly, including the MOS events, a search in the frequency range from $10^{-3}$ to $3$~Hz also did not uncover any significant modulations. In order to quantify the sensitivity of the conducted observations, we used Eq.~(5) in \citet{1987A&A...175..353B} to estimate limits on the pulsed fraction. At a $5\sigma$ significance, for EPIC-pn in the frequency range $10^{-3} - 13.5$~Hz we determined pulsed fraction limits of 13.1\% and 13.5\% for the two brightest sources \oeon\ and \oefo, and higher values of 15.4\%, 17\%, and 19.1\% for \ostt, \otfs, and \offf, respectively.

Next to short-term pulsations, the long-term flux evolution may also give insights into the nature of the sources. To this end, we used the HILIGT tool \citep{2022A&C....3800531S,2022A&C....3800529K} to obtain flux values and $3\sigma$ flux upper limits from the \xmm\ Slew Survey \citep{2008A&A...480..611S}, Swift/XRT \citep{2005SSRv..120..165B}, and the ROSAT PSPC All-Sky Survey and PSPC or HRI pointed observations \citep{1987SPIE..733..519P,2016A&A...588A.103B} to build long-term light curves in the 0.2 -- 12~keV band for all five candidates. Based on the best-fit spectral fit results (Table~\ref{tab_fitres}), we used the pre-defined HILIGT grid-point closest to the obtained fit parameters in the flux computation. The light curves were further expanded by including the flux measurement of the best-fit model and flux values from the four or five \eros\ sky scans that covered the positions of the targets. The resulting light curves are shown in Fig.~\ref{fig_long_term_lc}. In order to quantify the variability, we computed the FLUX\_VAR parameter, as exemplarily described in \citet{2020A&A...641A.137T}, from the \xmm, eRASS, and detected ROSAT survey flux values. With typical values $<2\sigma$, the light curves do not imply significant variability for any of the sources.

\begin{table*}
\caption{Photometric parameters\label{tab_optres}}
\centering
\scalebox{1.}{
\begin{tabular}{lrccccc}
\hline\hline
 & \ostt & \oeon & \oefo & \otfs & \offf\tablefootmark{(a)} \\
\hline
Magnitude zero-point (ZP)\tablefootmark{(b)} [mag]  &  28.30    &  28.18    &  28.18  &  28.18  &  28.18 \\
Extinction (E)\tablefootmark{(b)} [mag]             &  0.105    &  0.087    &  0.100  &  0.074  &  0.100   \\
Airmass (AM)                                        &  1.27     &  1.22     &  1.19   &  1.05   &  1.17   \\
FWHM [pixel]                                        &  2.64     &  3.26     &  3.25   &  2.56   &  3.46 \\
FWHM [\arcsec]                                      &  0.67     &  0.82     &  0.82   &  0.64   &  0.87 \\
$\sigma_{\rm sky}$                                  &  0.12     &  0.15     &  0.16   &  0.11   &   \\
Detection limit\tablefootmark{(c)} [$5\sigma$; mag] & 27.48     &  26.87    &  26.86  &  27.57  &  24.50 \\
X-ray-to-optical flux ratio [$5\sigma$]             & $>3000$   & $>5400$   & $>3100$ & $>3300$ & $200$ \\
Distance to nearest optical neighbour [\arcsec] & 2.43 & 3.65 & 3.53 & 2.70 & 0.14 \\
\hline
\end{tabular}}
\tablefoot{
\tablefoottext{a}{Optical magnitude of the likely counterpart given instead of a detection limit for \offf.}
\tablefoottext{b}{Nightly zero-point and extinction values applied from the FORS Absolute Photometry project (\url{https://archive.eso.org/qc1/qc1_cgi?action=qc1_browse_table&table=fors2_photometry}).}
\tablefoottext{c}{$5\sigma$ detection limit calculated as in \citet{2025A&A...694A.160K}}
}
\end{table*}


\subsection{Radio results for \oeon\ and \oefo}

We searched for radio pulsations, including both periodic and single-pulse signals, in the FAST observations of \oeon\ and \oefo. However, no radio signals were confirmed with a significance level of $\geq 8\sigma$. We can estimate an upper limit on the flux density of undetected pulsations with the radiometric equation \citep{2004hpa..book.....L}:
$$
\begin{aligned}
S_{\rm periodic} = \beta \displaystyle  \frac{(S/N) T_{\mathrm sys}}{G \sqrt{n_{\mathrm p} T_{\mathrm int} \Delta BW}}\sqrt{\frac{W_{\rm obs}}{P-W_{\rm obs}}} ,\quad
\end{aligned}
\hfill (1)
$$
where $\beta$ is the sensitivity degradation factor, $S/N$ is the threshold signal-to-noise ratio required for detection, $T_{\rm sys}$ is the system temperature, $G$ is the telescope gain, $n_{\rm p}$ is the number of polarisations, $T_{\rm int}$ is the integration time, $\Delta BW$ is the bandwidth in MHz, $W_{\rm obs}$ is the pulse width, and $P$ is the spin period. Here, the inserted parameters for the FAST observations are $\beta = 1$, $G = 16\,$~K\,Jy$^{-1}$, $T_{\rm sys}$ is $\sim$24~K, $n_{\rm p} = 2$, and $\Delta BW = 400\,{\rm MHz}$ \citep{2020RAA....20...64J}. For $W_{\rm obs}$ and $P$, we assumed $W_{\rm obs} = 0.1P$ and $S/N = 8$. The non-detection of signals sets $8\sigma$ flux density upper limits at 4.08\,µJy for J0819 and 2.72\,µJy for J0840.

For the single pulse search, the sensitivity limits can be estimated by (e.g. \citealt{2003ApJ...596.1142C,2009ApJ...702..692K}): 
$$
\begin{aligned}
S_{\rm single} &= 2\beta \frac{(S/N_{\mathrm{peak}}) T_{\mathrm{sys}}}{G \sqrt{n_{\mathrm{p}} W_{\rm obs} \Delta \mathrm{BW}}},
\end{aligned}
\hfill (2)
$$
where $S/N_{\rm peak}$ is the peak signal-to-noise ratio of a pulse; the other quantities are the same as in Eq.~(1). Single pulses are typically unstable and may show different widths \citep[e.g.][]{2014MNRAS.443.1463S}. Assuming $W_{\rm obs} = 1\,{\rm ms}$, we derive an upper limit on the single-pulse flux density of $\sim$26.8\,mJy for both targets.


\subsection{Limits on the optical emission}

\begin{figure*}[t]
\begin{minipage}[t]{0.33\textwidth}
\includegraphics[width=\linewidth]{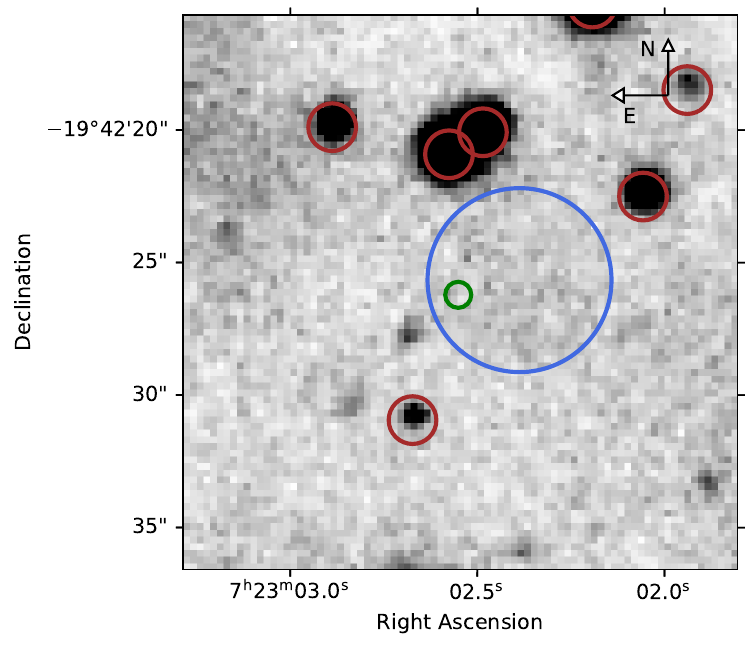}\\
\hspace*{1.5mm}\includegraphics[width=\linewidth]{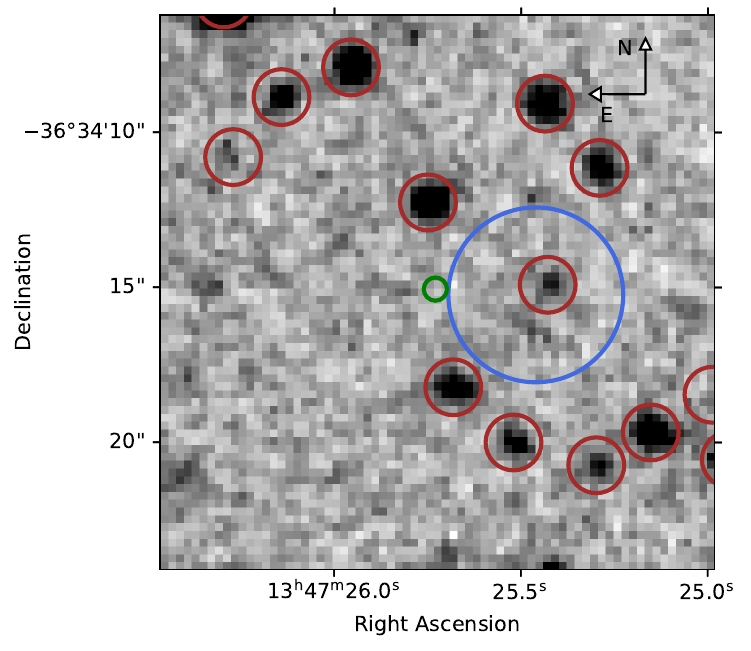}
\end{minipage}
\begin{minipage}[t]{0.33\textwidth}
\includegraphics[width=\linewidth]{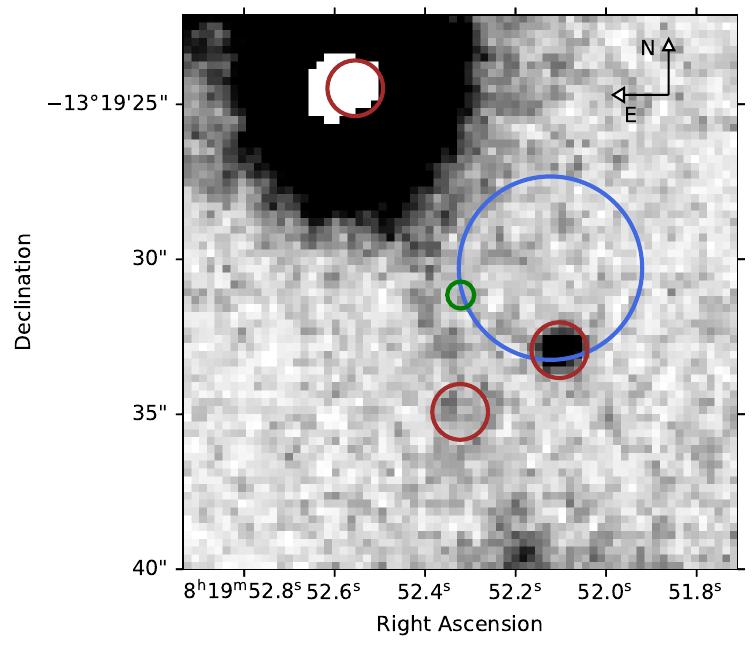}\\
\includegraphics[width=\linewidth]{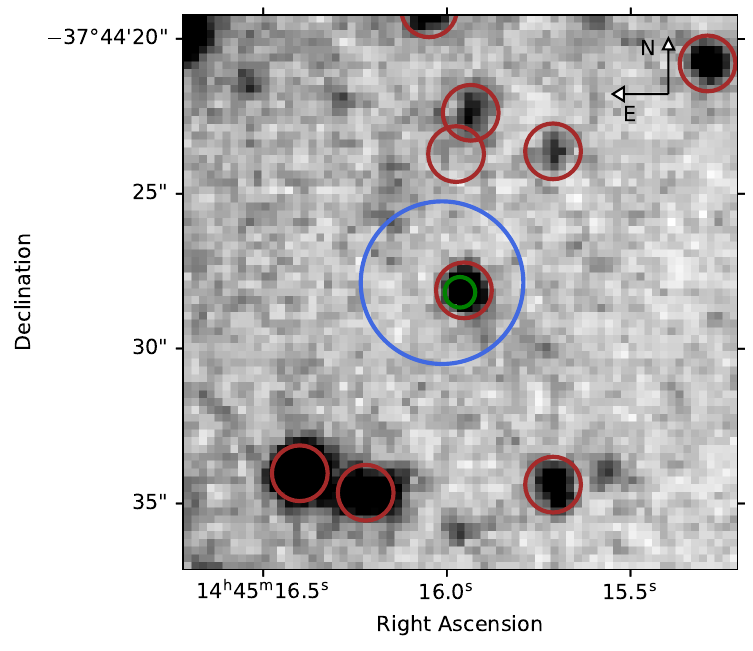}
\end{minipage}
\begin{minipage}[t]{0.33\textwidth}
\includegraphics[width=\linewidth]{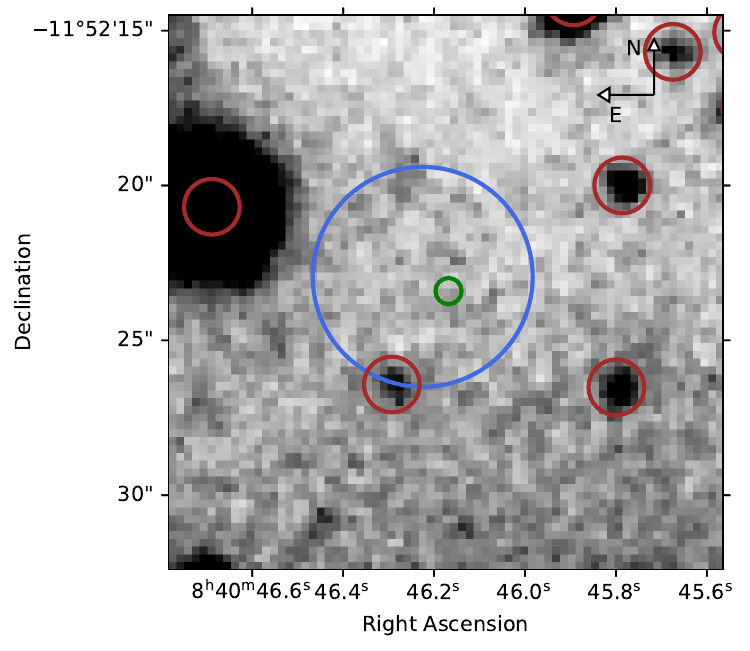}\\
\hspace*{1.mm}\includegraphics[width=\linewidth]{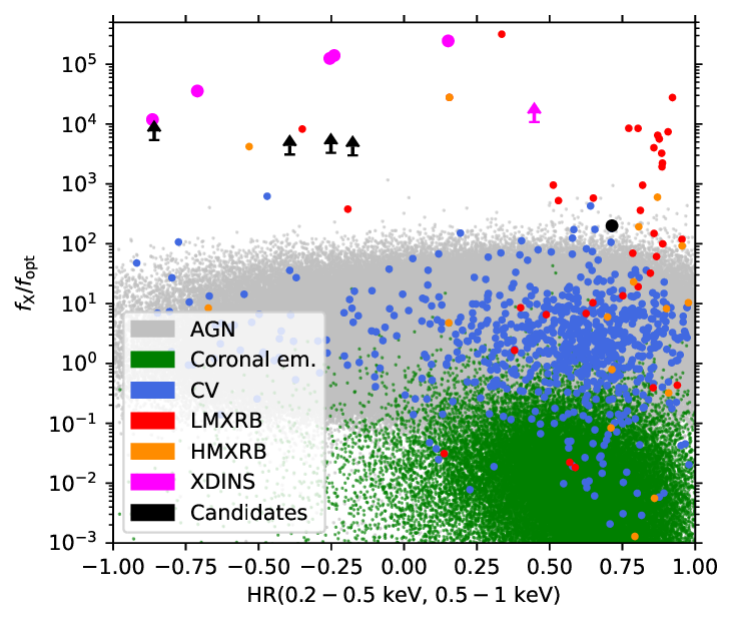}\\
\end{minipage}

\caption{\forst\ $R\_SPECIAL$ band finding charts of the fields containing the five INS candidates. We indicate the X-ray sky localisation from \xmm\ (green; 90\% confidence region; Table~\ref{tab_pos}), the X-ray sky position from \eros\ \citep[blue; 90\% confidence region;][]{2024A&A...687A.251K}, and the nearby field sources identified from a SExtractor run \citep[brown, arbitrary radii;][]{1996A&AS..117..393B}. (\textit{Bottom right:}) X-ray-to-optical flux ratio vs hardness ratio diagram, indicating the five candidates (black markers), the known XDINSs on the western Galactic hemisphere, including the recently discovered candidate \jotos\ \citep[magenta;][]{2024A&A...683A.164K}, as well as other soft X-ray emitting sources, namely AGNs \citep{2025arXiv250902842S}, coronal emitters \citep[green;][]{2024A&A...684A.121F}, CVs \citep[blue;][]{2003A&A...404..301R,2017ApJ...835...64G}, low-mass XRBs \citep[red;][]{2023A&A...675A.199A}, and high-mass XRBs \citep[orange;][]{2023A&A...677A.134N}.}
\label{fig_vlt_imgs}
\end{figure*}


All five candidates were observed with the \forst\ instrument at \eso\ in the $R\_SPECIAL$ band (Table~\ref{tab_obs}). We present the resulting images in Fig.~\ref{fig_vlt_imgs} along with the eROSITA and \xmm\ position. The deep optical imaging revealed a counterpart at $R\sim24.5$~mag (see Table~\ref{tab_optres}) within the error circle of the eROSITA and \xmm\ observations for \offf. We note that this counterpart is also included in the Legacy Survey DR10 release \citep{2019AJ....157..168D}, with $r=24.43$~mag, $i=23.72$~mag, and $z=22.65$~mag. However, it is undetected in the $g$ band. Consequently, the counterpart appears to possess a reddish colour. The remaining four candidates cannot be associated with any optical source in the deep imaging observations, as we find significantly ($>8\sigma$) separating angular distances of 2.4 -- 3.7\arcsec\ between the \xmm\ position and the nearest optical source.

In order to estimate the X-ray-to-optical flux ratios for the four candidates lacking optical counterparts, we determined the $5\sigma$ detection limits by applying the equation listed in the caption of Table~6 in \cite{2025A&A...694A.160K}. We present the photometric parameters used to infer the detection limit in Table~\ref{tab_optres}. The resulting magnitude limits, ranging from 26.9 -- 27.6~mag, imply lower limits on the X-ray-to-optical flux ratio of 3000 -- 5400. This indicates a compact nature for these sources (see X-ray-to-optical flux ratio diagram in Fig.~\ref{fig_vlt_imgs}). For \offf, the inferred magnitude of $\sim24.5$~mag in the R band implies an X-ray-to-optical flux ratio of $\sim200$ indicating a non-INS nature.


\section{Discussion\label{sec_disc}}

The five sources discussed here were originally selected from the eRASS on the premise of being promising thermally emitting INS candidates. The follow-up results presented above now allow for a more detailed discussion of the candidates' nature.

\begin{figure}[t]
\includegraphics[width=\linewidth]{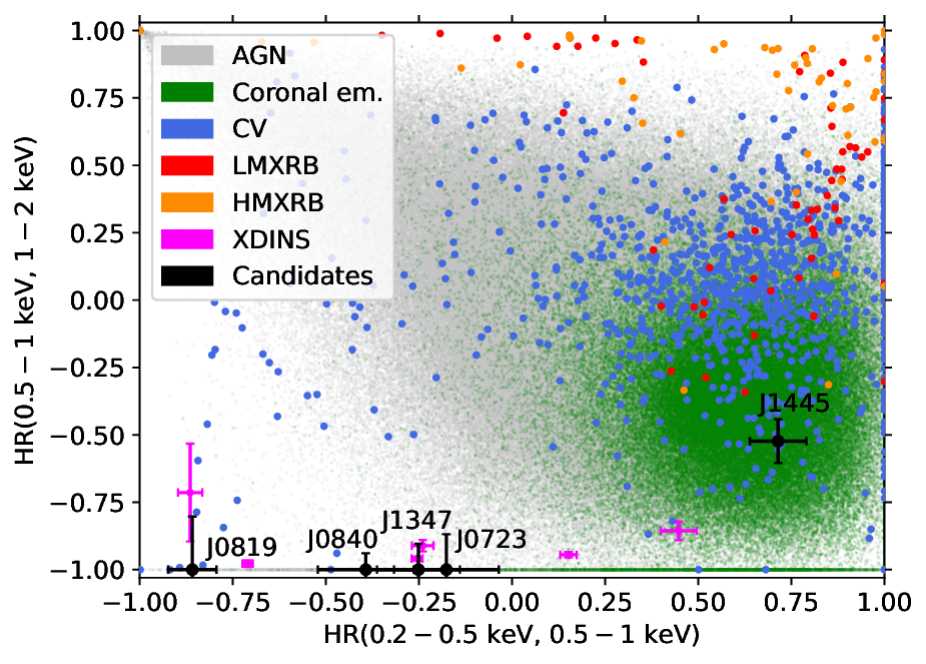}\\
\caption{Hardness-ratio diagram presenting the candidates (black), XDINSs including \jotos\ (magenta), and prevalent soft X-ray emitting source types such as AGNs (grey), coronal emitters (green), CVs (blue), low-mass XRBs (red), and high-mass XRBs (orange).}
\label{fig_hr_ratio}
\end{figure}

Optical counterparts (Fig.~\ref{fig_vlt_imgs}) could not be detected down to large X-ray-to-optical flux ratios ($>3000-5400$, Table~\ref{tab_optres}) in four instances (\ostt, \oeon, \oefo, \otfs). We also found the spectral continua of these sources to be well described with single absorbed \texttt{BB} or \texttt{NSA} components, indicating an overall thermal nature of their X-ray emission. Whereas based on X-ray-to-optical flux ratio alone an X-ray binary (XRB) nature may appear reasonable, the overall soft X-ray emission of these sources is more in line with the emission observed in predominantly thermally emitting INSs (see hardness ratio chart in Fig.~\ref{fig_hr_ratio}). Consequently, an INS nature can be favoured for these four sources.

The spectral modelling has shown that a single \texttt{BB} or \texttt{NSA} component is sufficient to fit the spectra well. Resolving heated polar caps or a non-uniform surface temperature distribution, as observed in most XDINSs and many RPPs \citep[e.g.][]{2019PASJ...71...17Y,2022A&A...661A..41S}, will require additional X-ray observations. The existence of non-thermal emission components originating in the INS magnetosphere and dominating the emission at energies above 1~keV could be excluded for these four sources to faint X-ray flux limits of $(1-5) \times 10^{-15}$~\fluxcgs\ ($3\sigma$ significance, 1 -- 12~keV band). Nevertheless, the resulting upper limits on the fraction of non-thermal to thermal emission ($\lesssim 1\%$) do not permit to fully rule-out magnetospheric emission components, as they are still shallow compared to the known population of thermally emitting INSs \citep{2005ApJ...623.1051D,2020ApJ...904...42D, 2022MNRAS.516.4932D}.

Based on the upper distance estimate from the best-fit $\nh$ value (Table~\ref{tab_fitres}), the sources not only are in line with Galactic X-ray emitters, but also imply thermal luminosities of the order of $\sim 10^{31}-10^{32}$~erg s$^{-1}$. Comparing these values to the luminosities of other known thermally emitting INSs \citep[e.g.][]{2020MNRAS.496.5052P}, we found them to be in accordance with an XDINS or RPP nature.

Most important to the characterisation and classification of INSs is the detection of pulsations, as a precise estimation of the neutron star spin properties allows us to constrain such basic physical parameters as characteristic age or dipolar magnetic field strength \citep{1969ApJ...157.1395O}. No significant periodic signals were discovered in any of the conducted X-ray observations (Fig.~\ref{fig_powspec_fits}) resulting in $3\sigma$ upper limits on the pulsed fraction of the four INSs at 13 -- 17\% ($10^{-3} - 13.5$~Hz band). At these limits, the non-detection of modulations is in line with the overall smooth X-ray pulsations observed in the known predominantly thermally emitting INS population. For example, only one to three sources out of the seven discovered \ros XDINSs possess modulations of similar strength or higher \citep[e.g.][]{2007Ap&SS.308..181H,2024ApJ...969...53B}. Additional observations are required to unveil the spin properties of the candidates in the X-ray regime.

Radio follow-up observations of \oeon\ and \oefo\ revealed no pulsations. This implies upper limits on the periodically pulsed radio flux density at $8\sigma$ significance of $4.08$\,µJy and $2.72$\,µJy for \oeon\ and \oefo, respectively. Single pulses are excluded at a flux density upper limit of $\sim$26.8\,mJy ($8\sigma$) for both sources. Radio-pulsars with flux densities $S_{1.4\mathrm{GHz}}<30$~µJy are often regarded as ‘radio-quiet’ \citep[e.g.][]{2023ApJ...958..191S}, implying that \oeon\ and \oefo\ must belong to the radio-faint population of pulsars. Their ‘radio-quiet’ nature is further supported by the fact that even with a very conservative distance upper limit of 2~kpc (higher than the $\nh$-based distance upper limits listed in Table~\ref{tab_fitres}) the resulting radio luminosity upper limits at 1.4~GHz are already fainter than those of 99\% of the pulsars with radio and distance estimates listed in the ATNF pulsar database \citep[version: 2.6.1;][]{2005AJ....129.1993M}.

In order to judge whether the current X-ray and radio limits imply unusual magnetospheric emission for \oeon\ and \oefo, we compared the obtained upper limits on the magnetospheric radio and X-ray emission to those of X-ray detected rotation-powered and millisecond pulsars (MSPs) catalogued in \citet{2025ApJ...981..100X}. Out of the $\sim 230$ pulsars, around 100 are also listed in the ATNF pulsar database and possess distance estimates and radio flux measurements at 1.4~GHz. From those, we computed radio to X-ray luminosity ratios and compared them to the upper limits on periodic modulations obtained for \oeon\ ($L_{1.4~\mathrm{GHz}}/L_{0.3-12~\mathrm{keV}}\leq 9 \times 10^{-15}$~GHz$^{-1}$) and \oefo\ ($L_{1.4~\mathrm{GHz}}/L_{0.3-12~\mathrm{keV}}\leq 4\times 10^{-15}$~GHz$^{-1}$). We found that 23 -- 30\% of the MSPs and RPPs possess even lower ratios, implying that the current X-ray and radio upper limits for \oeon\ and \oefo\ do not possess a particularly strong ratio between the undetected X-ray and radio magnetospheric emission components.

Magnetospheric emission can also manifest itself in the gamma-ray regime. We checked data release 4 of the fourth Fermi-LAT source catalogue \citep{2023arXiv230712546B} and the third Fermi-LAT catalogue of gamma-ray pulsars \citep{2023ApJ...958..191S} for possible counterparts. We did not find any matches for the five candidates, implying the absence of detectable gamma-ray emission at the current limits in these sources. While the predominantly thermal X-ray emission and the absence of detectable magnetospheric emission components in the four INSs may at first sight suggest similarities to the ‘radio-quiet’ XDINSs \citep[e.g.][]{2008AIPC..983..348K} and the recently discovered RPP \jzsfs\ \citep{2025A&A...694A.160K}, we emphasise that the current X-ray and radio limits for these four sources do not support claims of unusually weak magnetospheric emission. It can be noted, however, that the established radio-faintness of \oeon\ and \oefo\ suggests that INS searches at X-ray energies are important to complement the Galactic INS population, as these two sources would be easily missed in conventional radio pulsar surveys.

The long-term X-ray flux evolution of all candidates is in line with stable, non-variable emission. While there are sufficiently large gaps and often shallow upper limits in the X-ray coverage that could allow for significant variability in the past, the regular coverage with eROSITA from January 2020 to February 2022, along with the more recently obtained \xmm\ flux values, indicate stable X-ray emission. Together with effective temperatures $\lesssim 80$~eV in the four likely INSs, these properties are in agreement with middle-aged INS classes, such as XDINSs or RPPs. This is in contrast with younger INS classes, such as central compact objects \citep[CCOs;][]{2017JPhCS.932a2006D}, often possessing higher effective temperatures or displaying significant variability as in the magnetars \citep{2017ARA&A..55..261K}.

For \oefo\ and \otfs, X-ray spectral modelling required the inclusion of absorption line components at $300-400$~eV. These features are detected with high significance and arise regardless of the chosen background region, although an imperfect estimation of the background in the source region causing the lines cannot be fully refuted. X-ray absorption features have been observed in a variety of INS types \citep[e.g.][]{2003A&A...403L..19H,2013Natur.500..312T,2022A&A...661A..41S} and are generally thought to arise from cyclotron resonances of charged particles in the neutron star magnetosphere \citep{2019A&A...622A..61S} or transitions in surface atoms \citep[e.g.][]{2007Ap&SS.308..191V}. Assuming such an intrinsic origin, electron and proton cyclotron resonances would imply magnetic field strengths of $(3-5)\times 10^{10}$~G and $(5-10)\times 10^{13}$~G, respectively. The magnetic field strengths in the electron case are too weak for most RPPs and might imply that the electrons are located away from the surface in the magnetosphere. Such a scenario was exemplarily discussed for \tmzsfs\ in \citet{2018ApJ...869...97A}. The proton cyclotron case may imply a highly magnetised nature, for example that of a high-B pulsar or XDINS, or the existence of a multipolar field structure that locally causes a stronger magnetic field. Assuming that the electronic transitions in surface atoms cause these features, we applied Eq.~(2) from \citet{2003ApJ...599.1293H} to compute that hydrogen ionisation implies strong magnetic fields of the order of $10^{13}-10^{14}$~G, whereas transitions in higher-$Z$ elements correspond to lower field strengths \citep[see][and references therein]{2014PhyU...57..735P}. We note that the physical characterisation of the lines would also benefit from the determination of the pulsar spin properties, as this would allow us to compare the dipolar magnetic field strength from timing measurements to the field strengths implied from the possible line-forming mechanisms. Similarly, phase-resolved studies would shed light on the emission geometries.

The detection of an optical counterpart for \offf\ and its subsequent X-ray-to-optical flux ratio of $\sim 200$ indicates a non-INS nature. Located at the upper end of the contaminant cloud in the X-ray-to-optical flux ratio diagram in Fig.~\ref{fig_vlt_imgs}, \offf\ is in a region occupied by AGNs, CVs, and XRBs. Similarly, the thermal X-ray spectrum, with an effective temperature of $\sim210$~eV places \offf\ at the soft end of the AGN and XRB hardness ratio distribution (Fig.~\ref{fig_hr_ratio}). The optical and near-infrared (nIR) colours inferred from the Legacy Survey counterpart do not allow us to further constrain the source nature. They imply, however, that the optical/nIR emission is not the extension of the best-fit X-ray \texttt{BB} component. This is implied from the red colour of the optical/nIR counterpart and from the optical/nIR flux densities exceeding the extrapolated \texttt{BB} by a factor of $10^5-10^6$. An extragalactic nature of the source is not fully excluded, as, for example, the 99\% confidence interval of the best-fit $\nh$ parameter for a \texttt{BB} fit to the EPIC-pn observation allows for hydrogen column densities exceeding the Galactic value. We note that \offf\ cannot be associated with any galaxy in the Heraklion Extragalactic Catalogue \citep[HECATE;][]{2021MNRAS.506.1896K} nor any known Milky Way globular cluster.

Discussing the possible source types in more detail, under the assumption of an AGN nature, \offf\ would possess a comparatively large X-ray-to-optical flux ratio. The predominantly thermal spectrum also deviates from the \texttt{PL}-like X-ray spectra typically observed in AGNs \citep[e.g.][]{2010A&A...512A..58I}. In very luminous AGNs, a soft thermal excess can be observed, often modelled with \texttt{BB}s of effective temperatures of $100-200$~eV \citep[e.g.][]{2004MNRAS.349L...7G}. \citet{2020MNRAS.491..532G} report flux ratios in the range of $10^{-2} - 5$ between the soft-thermal excess and the non-thermal comptonised continuum. Assuming a ratio of 5 for \offf, its high-energy component (approximated by a power-law with a photon index of $\Gamma=2$) should possess a flux of $\sim10^{-14}$~\fluxcgs\ in the 3 -- 12~keV band. This is $4\sigma$ above the detection limit of the \xmm\ observation of \offf, indicating that such an excess should have been detected. We note, however, that a steeper non-thermal continuum would be undetectable in the present \xmm\ data (e.g. $\Gamma \gtrsim 2.3$ and $F_\mathrm{BB}/F_\mathrm{PL} = 5$ would result in a \texttt{PL} flux below the $3\sigma$ sensitivity limit at 3 -- 12~keV). Such a steeper non-thermal continuum may also be expected for the required large thermal-to-non-thermal flux ratio as such a correlation is proposed in \citet{2020MNRAS.491..532G}. Consequently, the current depth of the \xmm\ observation does not allow fully excluding the AGN nature; however, it is clear that \offf\ would need to be a rather extreme member of this class.

In many CV types (e.g. polars, intermediate polars, and luminous super-soft X-ray sources) the soft X-ray emission can be modelled well by \texttt{BB}-like emission components \citep[e.g.][]{2017PASP..129f2001M} with typical effective temperatures in the range of $\sim 10-100$~eV \citep[e.g.][]{1997ARA&A..35...69K,2020AdSpR..66.1209D, 2024A&A...690A.243S}. Assuming such a nature, \offf would be an unusually hot member of these source classes. Compared to the previously discussed AGN, CVs have been observed to possess even larger ratios between soft and hard components \citep[e.g. $\sim 0.1 - 100$ as reported in][]{2024A&A...690A.243S}. For this reason, the non-detection of a hard X-ray tail in the \xmm\ observation does not refute a CV nature. Assuming Galactic distances ($\lesssim 3$~kpc), \offf's luminosity ($\sim 10^{31}-10^{32}$~erg~s$^{-1}$) is in line with most CV types \citep[e.g.][]{2024A&A...690A.243S} but is too low for the luminous super-soft X-ray source class \citep[$10^{36}-10^{38}$~erg~s$^{-1}$;][]{1997ARA&A..35...69K}. The \texttt{BB}-like emission in CVs often implies radius values comparable to the size of white dwarfs \citep{2017PASP..129f2001M}. The obtained radius ($\sim 280$~m at 1~kpc distance; Table~\ref{tab_fitres}) is much too small for a Galactic CV source. We note that a simple bremsstrahlung model, often used to fit the X-ray continuum emission in CVs, can also model the spectrum of \offf\ well ($\chi^2_\nu(\nu) = 1.0(36)$, $\nh= 17.9^{+1.6}_{-1.5} \times 10^{20}$~cm$^{-2}$, $kT=437^{+22}_{-21}$~eV), even though the resulting plasma temperature is too low for most CVs exhibiting bremsstrahlung emission \citep{2017PASP..129f2001M}. APEC models with redshift values fixed to zero, on the other hand, do not provide good fits, as they either converge to unreasonably small abundance values or yield a poor fit statistic. Many CVs are also observed to show strongly modulated emission over a large frequency range \citep{2017PASP..129f2001M}. While not fully excluding a CV identification, the absence of significant variability in \offf\ is for this reason untypical. To conclude, given the present data, a CV nature appears unlikely for this source.

The emission of XRBs is highly dependent on the accretion rate, leading to typical luminosities of $10^{33}-10^{39}$~erg~s$^{-1}$ in accretion-dominated sources \citep[e.g.][]{2023hxga.book..120B,2023arXiv230802645F} and lower luminosities \citep[e.g. $10^{31}-10^{34}$~erg~s$^{-1}$;][]{2017JApA...38...49W} during quiescent low-accretion states. For \offf, the inferred luminosity of $\sim 10^{31}- 10^{32}$~erg~s$^{-1}$ would be more in line with low accretion rates and a quiescent XRB nature. This is further supported by the fact that the predominantly thermal X-ray emission of \offf\ originating from a small region (Table~\ref{tab_fitres}) is reminiscent of a neutron star hot spot. Such emission components can be detected in neutron star XRBs during quiescent states \citep[e.g.][]{2016MNRAS.463...78E}. On the other hand, XRBs are characterised by their X-ray outbursts and significant variability. At the time of writing, there are no hints towards a transient behaviour in \offf, but the observational baseline is still too short to use this fact to exclude an XRB nature, given that neutron star XRBs were already observed to remain in quiescence for multiple years \citep{2025ApJS..279...57H}. The optical emission in \offf\ may indicate the presence of a binary companion, a possible donor in an XRB scenario. For this reason and in the absence of significant accretion processes, the source may alternatively be identified as a detached binary pulsar. This would be more in line with stable X-ray emission and lack of non-thermal emission components. Binary companions to pulsars are most frequently observed in MSPs, where a significant fraction of them is located in such systems \citep{2017JApA...38...42M}. An MSP nature may also be supported by the X-ray emission properties of \offf, as similar hot spot-dominated spectra were observed in known thermally emitting MSPs such as \onon\ \citep{2017JApA...38...42M}. The fact that \offf's X-ray emission appears to originate from only a small part of its surface may lead to detectable modulations. The current limits, namely the pulsed fraction upper limit of 19\% and the relatively low time resolution of the \xmm\ observation, insensitive to the fast rotation periods observed in MSPs, provide only weak constraints on their presence. Consequently, additional X-ray follow-up is required to gain insights into the nature and properties of the possible neutron star. Similarly, an optical/nIR spectrum of \offf\ could prove very helpful, not only to explore alternative source classifications (e.g. AGNs), but also to further study the individual emission components in the source.

\section{Summary and conclusions\label{sec_concl}}
The conducted radio, optical, and X-ray follow-up observations confirm the INS nature of four sources (\ostt, \oeon, \oefo, and \otfs). Combining their stable and predominantly thermal X-ray emission with the exclusion of counterparts outside the X-ray regime, their emission properties agree mostly with those of intermediately aged INS classes such as XDINSs and RPPs. The detection of an optical counterpart, discovered for the remaining source (\offf), is most in line with an AGN or a binary pulsar nature, in either a detached or low-accretion state. While the conducted observations further constrain the possible source nature of the five candidates, they are still important targets for additional investigation. Thus, extended X-ray timing studies aimed at establishing neutron star spin periods are crucial to unambiguously place them among the population of Galactic INSs. For \offf, an optical/nIR spectrum could prove very valuable in breaking the degeneracy between the possible source types. It is, however, already clear that following the discoveries of \jzsfs\ and \jotos, the characterisation of eRASS-selected INS candidates continues to complement the known INS population with predominantly thermally emitting sources.


\begin{acknowledgements}
We thank the anonymous referee for helpful feedback and comments that improved this paper. This work was funded by the Deutsche Forschungsgemeinschaft (DFG, German Research Foundation) through grants Schw 536/38-1 and Schw 536/38-2, and by Deutsches Zentrum für Luft- und Raumfahrt (DLR) through grant Fkz 50\,OR\,2408.

AMP acknowledges the Innovation and Development Fund of Science and Technology of the Institute of Geochemistry, Chinese Academy of Sciences, the National Key Research and Development Program of China (Grant No. 2022YFF0503100), the Strategic Priority Research Program of the Chinese Academy of Sciences (Grant No. XDB 41000000), and the Key Research Program of the Chinese Academy of Sciences (Grant NO. KGFZD-145-23-15).

IT gratefully acknowledges the support by Deutsches Zentrum für Luft- und Raumfahrt (DLR) through grant 50\,OX\,2301.

This research has made use of data and/or software provided by the High Energy Astrophysics Science Archive Research Center (HEASARC), which is a service of the Astrophysics Science Division at NASA/GSFC.

This work has used the data from the Five-hundred-meter Aperture Spherical radio Telescope (FAST). FAST is a Chinese national mega-science facility, operated by the National Astronomical Observatories of Chinese Academy of Sciences (NAOC).

Based on observations made with ESO Telescopes at the La Silla Paranal Observatory under programme ID 111.259R.001

This work is based on data from eROSITA, the soft X-ray instrument aboard SRG, a joint Russian-German science mission supported by the Russian Space Agency (Roskosmos), in the interests of the Russian Academy of Sciences represented by its Space Research Institute (IKI), and the Deutsches Zentrum für Luft- und Raumfahrt (DLR). The SRG spacecraft was built by Lavochkin Association (NPOL) and its subcontractors, and is operated by NPOL with support from the Max Planck Institute for Extraterrestrial Physics (MPE).

The development and construction of the eROSITA X-ray instrument was led by MPE, with contributions from the Dr. Karl Remeis Observatory Bamberg \& ECAP (FAU Erlangen-Nuernberg), the University of Hamburg Observatory, the Leibniz Institute for Astrophysics Potsdam (AIP), and the Institute for Astronomy and Astrophysics of the University of Tübingen, with the support of DLR and the Max Planck Society. The Argelander Institute for Astronomy of the University of Bonn and the Ludwig Maximilians Universität Munich also participated in the science preparation for eROSITA.

This work made use of Astropy\footnote{http://www.astropy.org}: a community-developed core Python package and an ecosystem of tools and resources for astronomy \citep{astropy:2013, astropy:2018, astropy:2022}. This research made use of ccdproc, an Astropy package for image reduction \citep{matt_craig_2017_1069648}.

\end{acknowledgements}
\bibliographystyle{aa}
\bibliography{ref_list}
\end{document}